\documentclass[sigconf]{acmart}

\def\BibTeX{{\rm B\kern-.05em{\sc i\kern-.025em b}\kern-.08emT\kern-.1667em\lower.7ex\hbox{E}\kern-.125emX}}
\usepackage{diagbox}
\usepackage{multirow}
\usepackage{subfigure}

\begin{document}

\title{Selecting Reliable Blockchain Peers via Hybrid Blockchain Reliability Prediction}

\author{Peilin Zheng}
\affiliation{%
  \institution{Sun Yat-sen University}
  \city{Guangzhou}
  \country{China}
}
\email{zhengpl3@mail2.sysu.edu.cn}

\author{Zibin Zheng$^*$}
\affiliation{%
  \institution{Sun Yat-sen University}
  \city{Guangzhou}
  \country{China}
}
\email{zibinzheng@yeah.net}

\author{Liang Chen}
\affiliation{%
  \institution{Sun Yat-sen University}
  \city{Guangzhou}
  \country{China}
}
\email{jasonclx@gmail.com}

\begin{abstract}
Blockchain and blockchain-based decentralized applications are attracting increasing attentions recently.
In public blockchain systems, users usually connect to third-party peers or run a peer to join the P2P blockchain network.
However, connecting to unreliable blockchain peers will make users waste resources and even lose millions of dollars of cryptocurrencies.
In order to select the reliable blockchain peers, it is urgently needed to evaluate and predict the reliability of them.
Faced with this problem, we propose H-BRP, Hybrid Blockchain Reliability Prediction model to extract the blockchain reliability factors then make personalized prediction for each user.
Large-scale real-world experiments are conducted on 100 blockchain requesters and 200 blockchain peers.
The implement and dataset of 2,000,000 test cases are released.
The experimental results show that the proposed model obtains better accuracy than other approaches.

\end{abstract}


\maketitle
\section{Introduction}
Blockchain is firstly proposed by Bitcoin \cite{nakamoto2008bitcoin}. It consists of a continuously growing list of records, called blocks, which are linked and secured using cryptography. In a period of time, each peer in the P2P transaction network records the transactions and package them into a block to join the blockchain. The blockchain is maintained by all the peers in the P2P network through a consensus protocol. 
In Bitcoin-like blockchain systems, after receiving the previous block, the peer will try to calculate the hash for the next block as soon as possible to get the rewards, such as cryptocurrencies. This competition is so called \emph{mining} and the mining users are called \emph{miners}. In Bitcoin-like mining, if a user connects to unreliable peers that returns the wrong block or old block, the user will never gain the cryptocurrency reward. 

Blockchain-based decentralized applications (\emph{DApp}) have gained a lot of attentions from both industry and academia in recent years \cite{zheng2017overview}. Most DApp users do not run a blockchain peer by themselves, but interact with the third-party peers. However, this kind of third-party peers had been reported to be unreliable \cite{imtokenlate}, leading to bad user experience and even cryptocurrency lost by users' misoperation.

\begin{figure}
\centering
\includegraphics[width=3.5in]{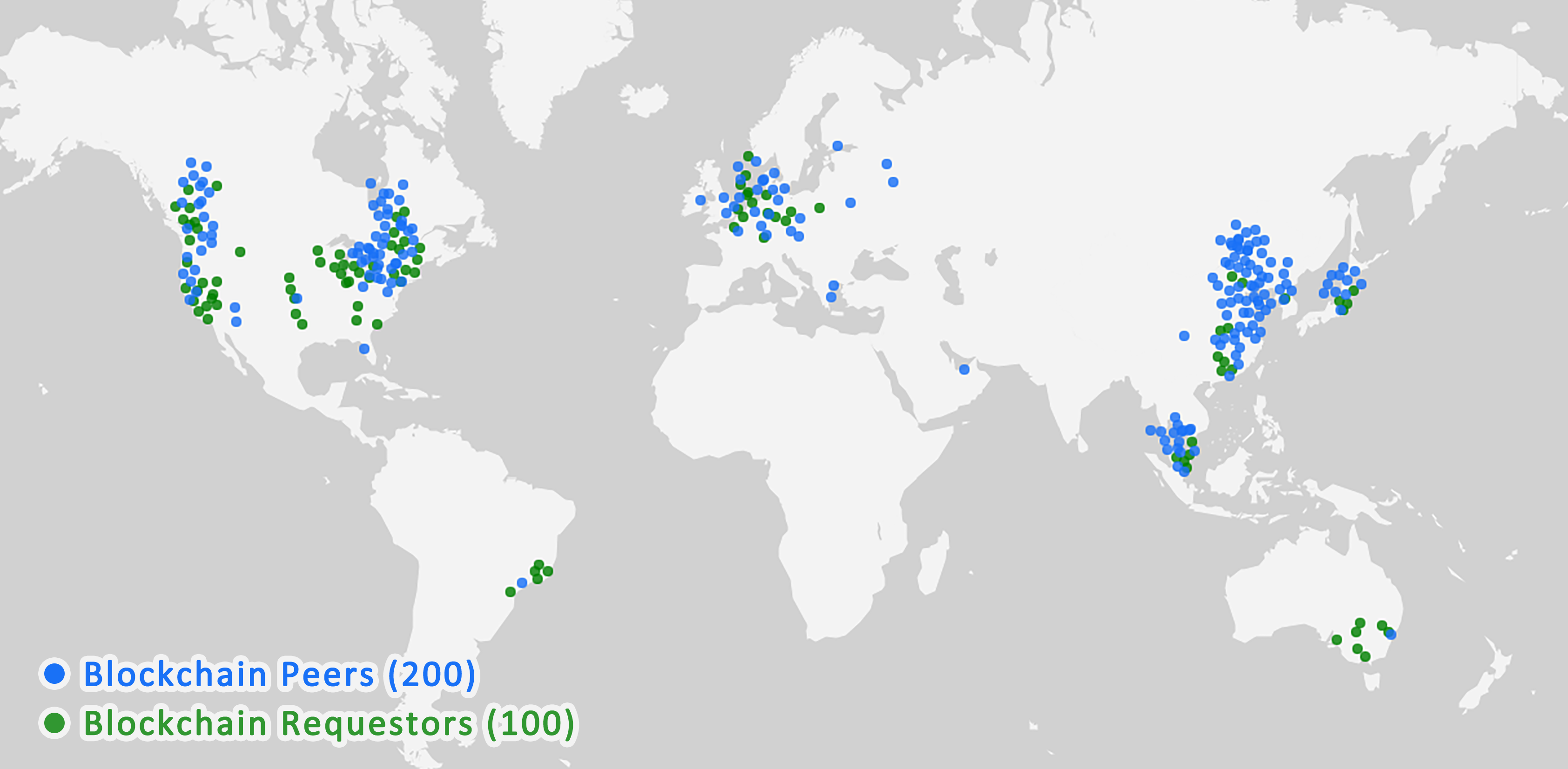}
\caption{Real-world Blockchain Peers and Requesters} \label{ipvis}
\end{figure}

Therefore, it is necessary for blockchain users to select the more reliable peers. The effect of selecting the peers can be summarized as two folds:
\textbf{(1) For blockchain mining:} The blockchain users' mining profit is proportional to reliability of the peers connected to it.
\textbf{(2) For blockchain-based application users:} The reliability of the peers determines the correctness and delay of transactions. Selecting reliable peers will help reduce the delay and avoid cryptocurrencies lost by repeated transactions. Thus there is great economic benefit and urgency to select reliable blockchain peers.
There are more than 20,000 blockchain peers at the same time in the real-world.
But single user cannot connect to all the peers in the meanwhile to evaluate their reliability so that the user need to predict the reliability.

There are some difficulties of blockchain reliability prediction 
As for blockchain reliability, Zheng et al.\cite{zheng2018detailed} and Dinn et al. \cite{dinh2017blockbench} propose the ways to evaluate the availability and performance of blockchain systems.
However, these methods enable only the owner of the peer to know the reliability, thus do not work for other users.
And, since the network situation is different for each user, the observed reliability of the same peer could be different for different users, which will be shown in Section~\ref{rq1}.
To attack this challenge, a personalized reliability prediction method is needed.

In this paper, we propose a hybrid collaborative reliability prediction model for blockchain systems, called \emph{H-BRP}. 
H-BRP does not predict the success rate of the blockchain peers directly. The main idea of H-BRP is to extract blockchain-related factors from the request history (e.g., block hash, block height). Then it uses the relationship between similar blockchain users and peers to do the collaborative prediction with hybrid linear regression. In this way, H-BRP obtains personalized prediction results for different users with higher accuracy than other approaches, as the real-world experiment shows. As shown in Figure~\ref{ipvis}, we deploy 100 blockchain requesters to evaluate and predict the reliability of 200 real-world blockchain peers, showing the feasibility and effectiveness of the model.

In summary, the main contributions of this paper are summarized as follows:

\begin{itemize}
\item We propose H-BRP, Hybrid Blockchain Reliability Prediction model for blockchain systems. This model can extract blockchain factors related to reliability. And, it uses the relationship between similar users and peers for personalized prediction.
\item We conduct real-world experiment with 2,000,000 test cases from 100 requesters to 200 blockchain peers as shown in Figure \ref{ipvis}. The results show the effectiveness of the proposed model. The implement and dataset will be open-source.
\end{itemize}

The rest of the paper is organized as follows. Section \ref{Basic Concepts} introduces the basic concepts of blockchain systems. Section \ref{Motivation} describes the motivating example of reliability prediction of blockchain systems. Section \ref{Methodology} proposes the details of the Hybrid Blockchain Reliability Prediction model, including the data processing, training and prediction. Section \ref{Implement and Experiment} introduces the implement of H-BRP and the experiment results. Section \ref{Related Work and Discussion} provides the related work and discussion about blockchain reliability. Section \ref{Conclusion and Future Work} concludes the paper and gives the future work.

\section{Basic Concepts} \label{Basic Concepts}
This section introduces the basic concepts of the blockchain and decentralize application.

\begin{figure}
\centering
\includegraphics[width=3.5in]{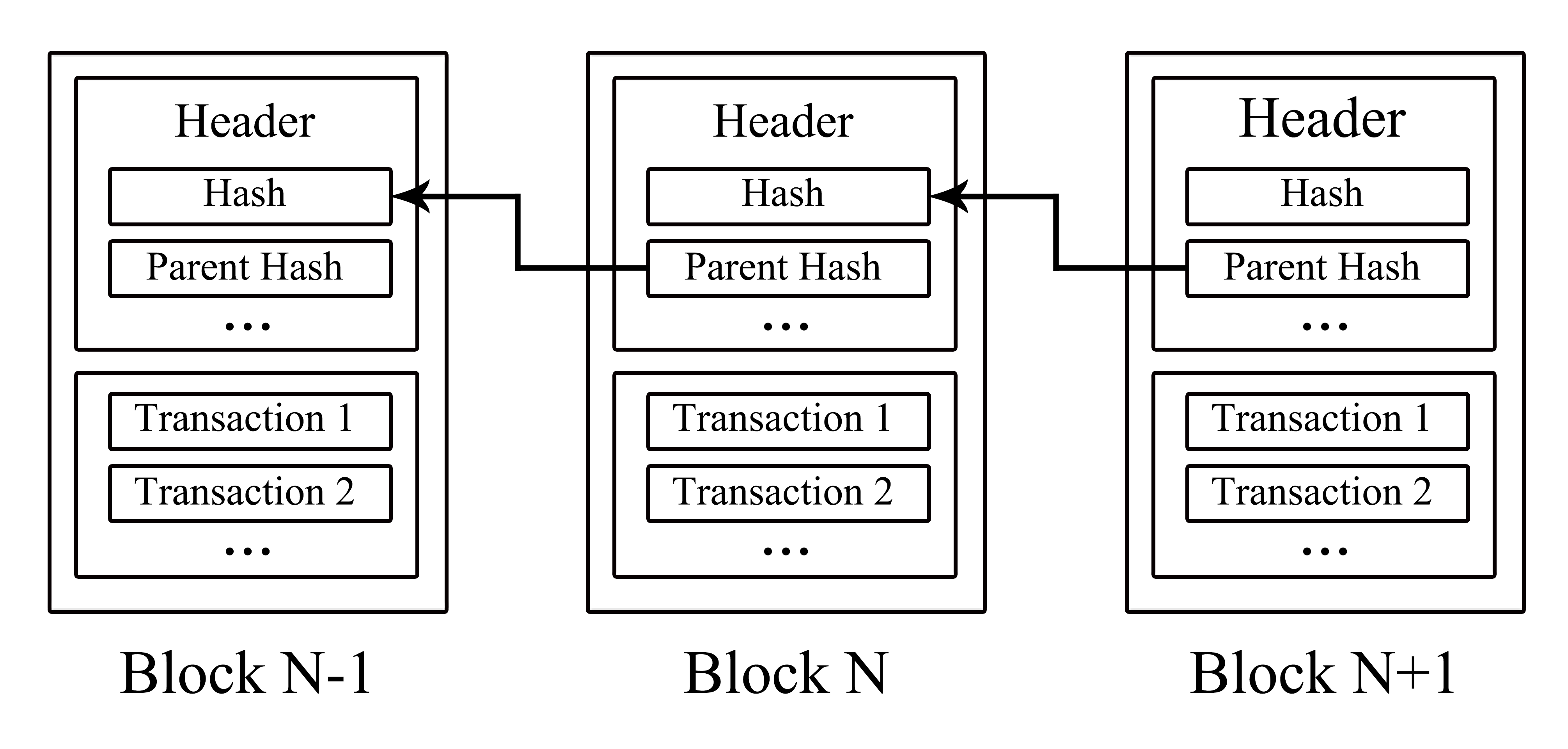}
\caption{Data Structure of Blockchain\cite{zheng2018detailed}} \label{blockchainview}
\end{figure}

In a narrow sense, the blockchain is a kind of data structure. The concept of the blockchain was firstly proposed as the underlying storage for peer-to-peer payments in Bitcoin\cite{nakamoto2008bitcoin}. As shown in Figure \ref{blockchainview}, every block contains the transactions in a period of time. Then every block is joint to a chain-like data structure named blockchain. Each peer in the peer-to-peer network maintains a blockchain by itself. And they keep it the same with each other via consensus protocols. Each block has a hash value of itself and this hash value is contained in the next block to make it tamper-resistant and traceable. 

\begin{figure}
\centering
\includegraphics[width=3in]{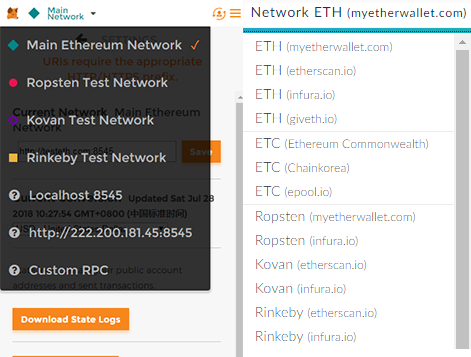}
\caption{Different Thrid-party Blockchain Peers in DApps} \label{blockchainapi}
\end{figure}

In a wide sense, the blockchain can be regarded as a new kind of distributed system. The basic concepts of blockchain are listed as follows:
\begin{itemize}
\item {Transaction: } A transaction represents a message to change the ledger, such as transferring cryptocurrencies. If someone wants to send Bitcoin to others, he should broadcast the transaction to the p2p network.

\item {Block: } Block is a data package consists of the transactions in a period of time.
Each block has a hash value of it self \cite{nakamoto2008bitcoin,buterin2013ethereum,wood2014ethereum} so that the hash value can be used to check the authenticity of the block.

\item {Chain: } The chain consists of all the blocks that are linked by their hash. In the Bitcoin blockchain\cite{nakamoto2008bitcoin}, every block is generated after the previous one so that they record the hash of the previous block. This chain-like structure is shown in Figure \ref{blockchainview}.

\item {Mining: } In public blockchain systems, after receive the previous block, the peers will try to find out a nonce for the next block to get the rewards, such as Bitcoin.
This process is so called \emph{mining} and the peers are called \emph{miners}. 

\item {Decentralized Application: } Blockchain-based decentralized applications (\emph{DApp}) use blockchain as the underlying technology \cite{what3,what2,YourfirstDapp,taipei,Stranford}.
Most DApp users connect to third-party peers to get the blockchain data. Figure~\ref{blockchainapi} shows different third-party blockchain peers in DApps.
If the user connect to an unreliable peer, the DApp would not work.

\end{itemize}

In summary, blockchian is a growing chain-like data structure maintained by peers in P2P network.
Each peer in the P2P network wants to get the latest (or so-called highest) correct block.
Therefore, a blockchain peer is reliable if it can return the latest block for the requesters.

\begin{figure}
\centering
\includegraphics[width=3.2in]{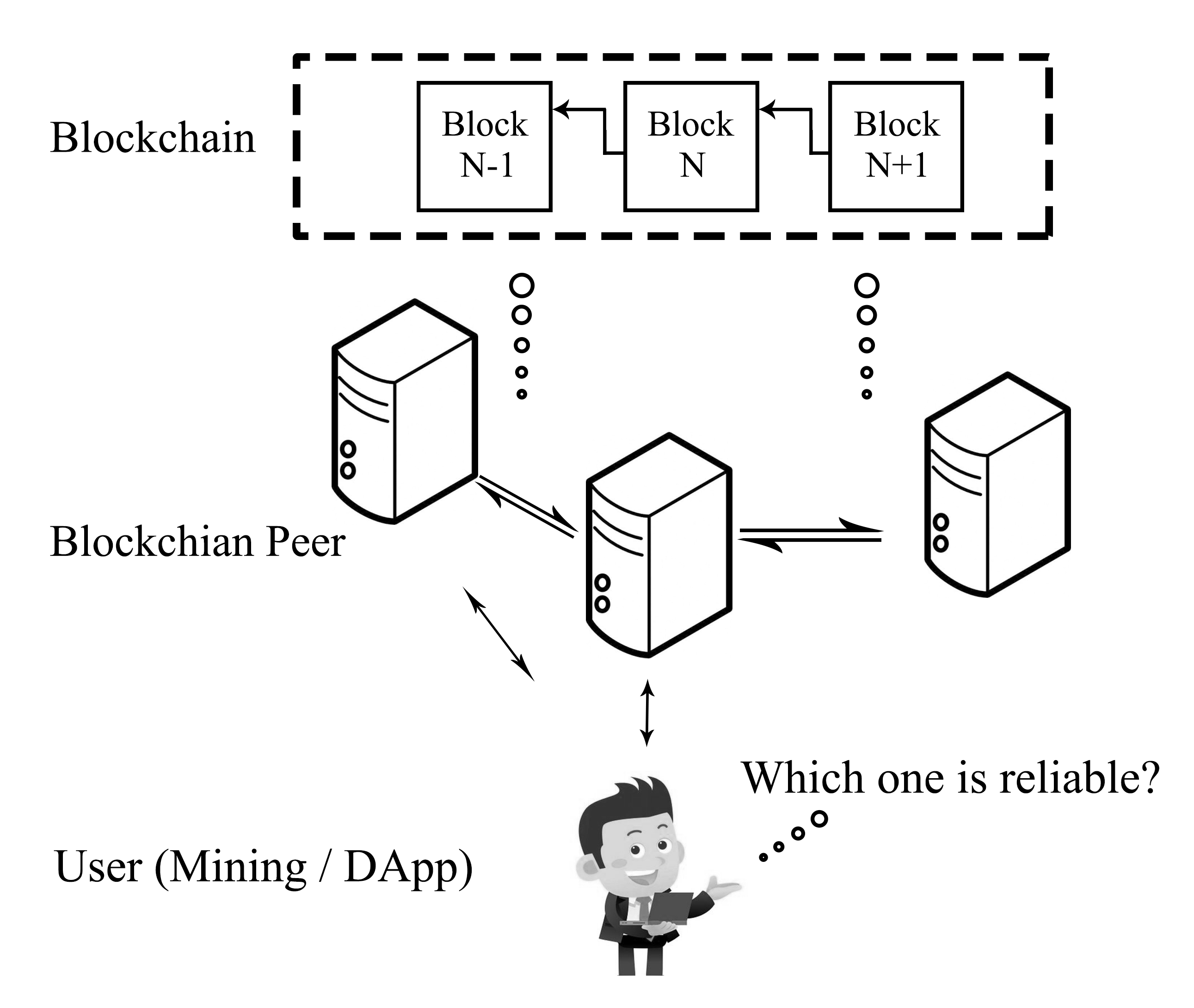}
\caption{Motivating Example of Blockchain Reliability Prediction} \label{arc3}
\end{figure}

\section{Motivating Example} \label{Motivation}
In this section, a motivating example of blockchain reliability prediction is given.
Blockchain users can choose either to run blockchain peers by themselves or use the third-party peers to interact with the blockchain systems.
As shown in Figure \ref{arc3}, no matter the user run a peer by himself or not, he should select some blockchain peers to connect to.
The challenge is that, there are more than 20,000 peers online at the same time, and he cannot test all of them to see which one is reliable for him.
The influence of the reliability of the blockchain peers can be divided into two folds: for blockchain mining and for blockchain-based application user.

\subsection{For Blockchain Mining}
The premise of blockchain mining is that the user should get the latest previous block.
If the user connects to the peers with low reliability, he cannot get the previous block in time.
Then the user will waste the computing resources in computing at the wrong block without any rewards of cryptocurrencies.
To improve the block synchronization speed and economic benefit, it is vital to evaluate and predict the reliability of the blockchain peers.

\subsection{For Blockchain-based Application User}
Assumed that the user in Figure~\ref{arc3} is like most blockchain-based application users that he connect to third-party blockchain peers.
Then he need to select the most reliable one since unreliable peer will cause consequences.
For example, one of the most famous wallet of cryptocurrencies called imToken had been reported to fail to sync with the Ethereum network \cite{imtokenlate}.
At that moment, the users think wrongly that their transactions are not confirmed by the network, then they send other repeated transactions again and again, which causes the loss of their money.

Thus it is really important for the blockchain users to know which blockchain peer is more reliable to avoid the loss of cryptocurrencies and improve the experience with the blockchain.

And, in reality, a user cannot connect to all the blockchain peers in the meanwhile, which means that there are lots of peer of unknown reliability.
Thus it is necessary to predict the unknown reliability of blockchain peers.

In summary, the motivations of this paper are to evaluate and predict the reliability of blockchain peers and help select the reliable peers to improve the blockchain synchronization speed, avoid loss of money, and improve the experience of blockchain.

\section{Methodology} \label{Methodology}

In this section, the methodology of Hybrid Blockchain Reliability Prediction model will be introduced, including the architecture and the steps in details.

\begin{figure}
\centering
\includegraphics[width=3.4in]{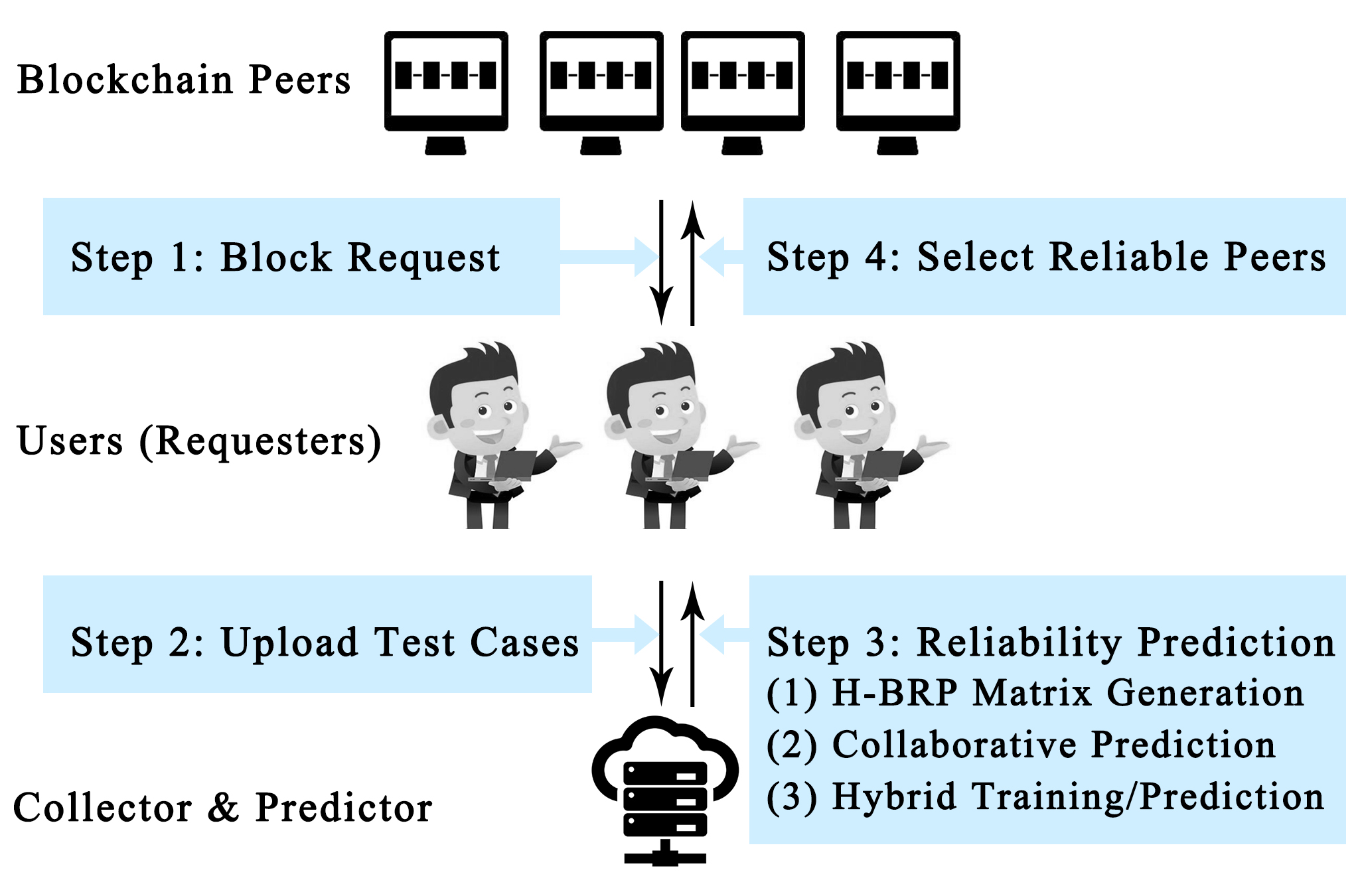}
\caption{Architecture of the Blockchain Reliability Prediction} \label{brt1}
\end{figure}

\subsection{Architecture}

The architecture of the blockchain reliability prediction is shown in Figure \ref{brt1}. It consists of 3 major roles as follows:
\begin{itemize}
\item \textbf{Blockchain Peers: } Blockchain peers are the nodes which maintaining the blockchain in the P2P network.
\item \textbf{Blockchain Requesters: } Requesters are the nodes installed with the H-BRP data collecting program, which will randomly request the blockchain data from some of the peers in a period of time. The requesters can be considered as the users in the blockchain systems.
\item \textbf{Data Collector \& Predictor: } Data collector is a central server that collecting all the test cases from the blockchain requesters. Then the data will be used to evaluate and predict the reliability of the blockchain systems.
\end{itemize}

The main idea of this architecture is that users can contribute their blockchain request history to a central data collector. Then the collector will summarize the historical data and do personalized reliability prediction for each user and peer.
As shown in Figure \ref{brt1}, there are 4 steps of Hybrid Blockchain Reliability Prediction: \emph{Block Request Testing}, \emph{Upload Test Cases}, \emph{Reliability Prediction}, and \emph{Select Reliable Peers}.

\subsubsection{\textbf{Block Request Testing}} \quad\\
In a period of time, a requester has more than one blockchain peer as candidate to connect to. Before knowing the candidates are reliable or not, the requester needs to connect to them. However, limited by the network conditions, the requester cannot request all the candidates in the meantime. Therefore, H-BRP proposes random batch block request testing for blockchain peers.

The random batch request testing include several stages as follows:
\begin{enumerate}
\item Given a batch size \textbf{\emph{n}}, and the time period of \textbf{\emph{t}} seconds,
\item For each time period, each requester selects \textbf{\emph{n}} candidates from the list randomly. 
\item Each requester requests the latest block from the selected candidates, parses it and records the height and hash to the local storage.
\end{enumerate}

\subsubsection{\textbf{Upload Test Cases}} \quad\\
After above stages, a requester achieves the raw data in a tuple of \emph{$<$ClientIP, BatchTime, PeerIP, StartTime, EndTime, Height, BlockHash$>$}.
The example is shown in Table \ref{rbbrt}.
In particular, H-BRP records the BatchTime to compare the test cases that are sent at the same time to see which one returns the higher block. H-BRP also records the StartTime and EndTime to see how long the round-trip time is during every test case. And, H-BRP records the Height and BlockHash in order to backtrack that whether the peer returns a correct block.

\begin{table*}
\caption{Example of Random Batch Block Request Testing}\label{rbbrt}
\centering
\begin{tabular}{|lllllll|}
\hline
{\bf RequesterIP} & {\bf BatchTime} & {\bf PeerIP} & {\bf StartTime} & {\bf EndTime}& {\bf Height} & {\bf BlockHash} \\ \hline
103.49.160.131 & 1532328744 & 167.99.208.120 & 1532328744 & 1532328744 & 6232293 & 0xa8b2b... \\ \hline
103.49.160.131 & 1532328744 & 219.117.201.187 & 1532328744 & 1532328744 & null & null \\ \hline
103.49.160.131 & 1532328744 & 116.62.100.69 & 1532328744 & 1532328744 & 5936957 & 0x073d4... \\ \hline
103.49.160.131 & 1532328744 & 147.75.80.165 & 1532328744 & 1532328745 & 6014476 & 0x7793a... \\ \hline
103.49.160.131 & 1532328744 & 47.75.9.16 & 1532328749 & 1532328749 & 6013794 & 0x8050b... \\ \hline
...&...&...&...&...&...&...\\ \hline
\end{tabular}
\end{table*}

When there are enough test cases, the requester can choose to upload the test cases to the collector.
The more test cases that the requesters contribute to the collector, the more accurate reliability prediction will be done.

\subsubsection{\textbf{Reliability Prediction}} \quad\\

After receiving enough test cases from users, the collector/predictor can choose different predicting model to do reliability prediction for the users and peers.
Reliability prediction is the key step in the whole architecture.
As shown in Figure~\ref{brt1}, H-BRP model includes three substeps: \emph{H-BRP Matrix Generation},\emph{Collaborative Prediction}, and \emph{Hybrid Training/Prediction}.
The detailed model and implement will be proposed in the next subsection. Here are the main ideas of it.

\begin{itemize}
\item \emph{H-BRP Matrix Generation:}
This substep can be regarded as the data preprocessing. H-BRP proposes some factors that are related to blockchain reliability. It transfers the data from a list of test cases into some metrics. In this substep, each factor is extracted into a requester-peer matrix.
\item \emph{Collaborative Prediction:}
Since blockchain peer shows different network delay to different users, the reliability observed by different users could be different. The case study in Section~\ref{Implement and Experiment} will show this difference.
Therefore, it is necessary for the model to do personalized prediction for different users.
To attack this problem, this substep is to find out similar blockchain users or peers, and then predict the unknown reliability factors for them.
\item \emph{Hybrid Training/Prediction:}
H-BRP assumed that there is a mapping between the reliability and factors extracted in previous substeps.
Thus the reliability can be predicted based on the prediction of the related factors.
In this substep, H-BRP first trains a linear regression model using the known reliability and factors.
After that, H-BRP uses this model and the predicted factors to do reliability prediction.
\end{itemize}

Sincerely, directly predicting the reliability without extracting the factors could be chosen. But direct reliability prediction will make it lose lots of valuable information from the source data.
That is why H-BRP extracts the factors from the test cases and do collaborative prediction by finding similar blockchain users/peers. 

In summary, the key idea is to maximize the use of available information, such as blockchain features and users' similarity.
The detailed model will be propose in next subsection.

\subsubsection{\textbf{Select Reliable Peers}} \quad\\
Based on the personalized reliability prediction result, the users can choose the blockchain peers with more reliability. For blockchain-based application users, the most reliable peer should be chosen. As for blockchain miners, they can choose top K peers ranked by predicted reliability.

\subsection{Hybrid Block Reliability Prediction Model} \label{Hybrid Block Reliability Prediction Model}

\begin{figure*}
\centering
\includegraphics[width=5.6in]{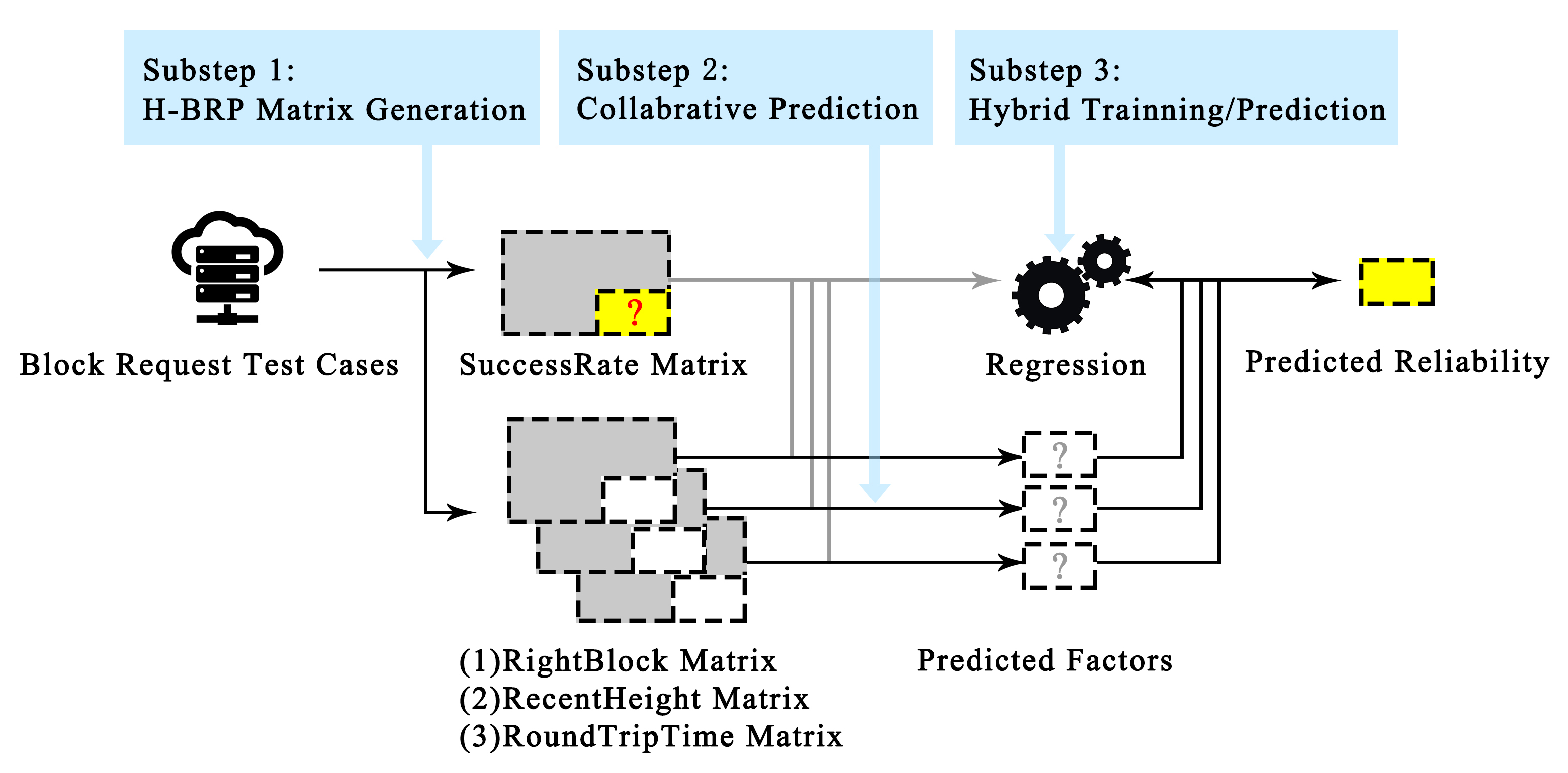}
\caption{Details of Hybrid Block Reliability Prediction Model} \label{brt}
\end{figure*}

In this subsection, the details of Hybrid Block Reliability Prediction Model will be described, as shown in Figure~\ref{brt}.

\subsubsection{\textbf{Blockchain Factor Matrix Generation}} \quad\\
After finishing the block request testing, the data collector use the request data to generate the Blockchain Factor Matrix.

First, Set up a blocks-tolerance value as \textbf{\emph{MaxBlockBack}} to represent the max tolerance for block backwardness of the peer in the blockchain.
Then set up a time-tolerance value as \textbf{\emph{MaxRTT}} to represent the max round-trip time for the peer.

The Success Rate Matrix is generated as follow:

For each requester \textbf{\emph{R$_i$}} and peer \textbf{\emph{P$_j$}} , set up a success counters for reliable requests as \textbf{\emph{SuccessRequest$_i,_j$}} and a failure counters as \textbf{\emph{FailureRequest$_i,_j$}}.

Next backtrack each batch of block requests to the peer, the peer responses successfully if and only if it:
\begin{enumerate}
\item \textbf{Returns right block: } The block hash is right in the corresponding block height on the main blockchain.
\item \textbf{Returns recent block height: } The block height subtracted from the highest one in the batch is no more than \emph{MaxBlockBack}. If \emph{MaxBlockBack} is set to 0, it requires the peer is reliable only when it returns the highest block in the batch.
\item \textbf{Returns in time: } The round-trip time of the request to the peer is no more than \emph{MaxRTT}.
\end{enumerate}

If the blockchain peer \emph{P$_j$} responses successfully in a batch, then count it into \emph{SuccessRequest$_i,_j$}, otherwise into \emph{FailureRequest$_i,_j$}.
Then the success rate of requester \emph{R$_i$} to peer \emph{P$_j$} can be calculated by :
\begin{equation}
TotalRequest_{i,j} = SuccessRequest_{i,j}+FailureRequest_{i,j}
\end{equation}
\begin{equation}
SuccessRate_{i,j} = \frac{SuccessRequest_{i,j}}{TotalRequest_{i,j}}
\end{equation}

After that, a matrix of success rate is achieved. As shown in Figure \ref{brt}, the gray area is the known success rates, while the yellow area is the unknown success rates, which is needed to predict. Some research in service computing use the success rate or failure rate to predict the unknown entries in the matrix to predict the reliability of service.
However, in blockchain reliability, this would lose some information from the source data because it do not take blockchain factors into account. Therefore, H-BRP generates three matrix corresponding to the above three blockchain related factors:

\begin{enumerate}
\item \textbf{Right Block Matrix: }

Right Block Matrix reflects on the rate at which the \emph{P$_j$} returns the correct block to \emph{R$_i$}. It can be generated by the following equation:
\begin{equation}
RightBlock_{i,j} = \frac{RightBlockRequest_{i,j}}{TotalRequest_{i,j}}
\end{equation}
where \emph{$RightBlockRequest_{i,j}$} is the counter of the requests that return the right block from \emph{P$_j$} to \emph{R$_i$}.

\item \textbf{Recent Height Matrix: }

Recent Height Matrix reflects on the rate at which the \emph{P$_j$} returns the recent height to \emph{R$_i$}. It can be generated by the following equation:
\begin{equation}
RecentHeight_{i,j} = \frac{RecentHeightRequest_{i,j}}{TotalRequest_{i,j}}
\end{equation}
where \emph{$RecentHeightRequest_{i,j}$} is the counter of the requests that return the recent height from \emph{P$_j$} to \emph{R$_i$}.

\item \textbf{Round-trip Time Matrix: }

Round-trip Time Matrix reflects on the average round-trip time of the block requests between \emph{P$_j$} and \emph{R$_i$}. It can be generated by the following equation:
\begin{equation}
RoundTripTime_{i,j} = \frac{\sum_{k} RTT_{i,j,k}}{TotalRequest_{i,j}}
\end{equation}
where \emph{$RTT_{i,j,k}$} is the round-trip time of the request from \emph{R$_i$} to \emph{P$_j$} in batch \emph{k}.
\end{enumerate}

In summary, in this phase, H-BRP generates one Success Rate matrix and three blockchain related factor matrices.
The main idea of the matrix generation is to extract more information related to blockchain in the source data. Thus the prediction using this data will be more accurate.

\subsubsection{\textbf{Collaborative Prediction}} \quad\\
After generating the matrices, the RightBlock Matrix, RecentHeight Matrix and RoundTripTime Matrix will be used into three collaborative filtering models.
The target is to predict the missing value in the blank of the matrix. As shown in Figure \ref{brt}, the target in this step is to predict the factors.
It is assumed that the three events (right block, recent height, and in time) corresponding to the matrix are independent. Thus every factor matrix will be predicted through the following phases independently.

\textbf{(1)Similarity Calculation}

In each matrix, H-BRP employ PCC to calculate the similarity between Blockchain Peers \emph{P$_i$} and \emph{P$_j$} by using: 
\begin{equation}
\label{eq-itempcc}
Sim(i,j) = \frac{\displaystyle\sum_{r \in R_i \cap R_j}{(m_{r,i} - \overline{m_i})}{(m_{r,j} - \overline{m_j})}}{\displaystyle\sqrt{\sum_{r \in R_i \cap
R_j}(m_{r,i} - \overline{m_i})^2}{\sqrt{\sum_{r \in R_i \cap R_j}(m_{r,j} - \overline{m_j})^2}}}, 
\end{equation}
where $R_i \cap R_j$ is a set of blockchain requesters that connected to both the blockchain peers $i$ and $j$, and $\overline{m_i}$ is the average value of the vector $i$ in the matrix

\textbf{(2)Similar Blockchain Peer Selection}

After calculating the similarity values between the peers, a set of similar peers can be identified by setting a parameter \emph{k} to select Top-k peers as similar peers to one specific peer.

To predict a missing factor entry $m_{r,i}$ in the factor matrix, a set of similar blockchain peers $SimPeers(i)$ with the blockchain peer \emph{P$_i$} can be identified by:

\begin{equation}
SimPeers(i) = \{k|Sim(i,k) \ge Sim_k, Sim(i,k) >0, k \neq i \},
\end{equation} where $Sim_k$ is the $k^{th}$ largest PCC value with blockchain peer \emph{P$_i$} and $Sim(k,i)$ can be computed by Equation~\eqref{eq-itempcc}.

\textbf{(3)Unknown Factor Prediction}

Employing the similar blockchain peers $SimPeers(i)$, H-BRP adopts item-based approaches~\cite{sarwar:item-based} (named as \emph{IPCC}) to predict the missing value $m_{r,i}$ by:
\begin{equation}
\label{eq-itempre}
m_{r,i} = \overline{m_i} + \sum_{k\in SimPeers(i)} w_k \times (m_{r,k}-\overline{m_k}),
\end{equation}
where $\overline{m_i}$ and $\overline{m_k}$ are average value of the blockchain peer $i$ and $k$
observed by different requesters, respectively, and $w_k$ is the significant weight of the similar blockchain peer $k$, which defined as:
\begin{equation}
w_k = \frac{Sim(i,k)}{\sum_{j\in SimPeers(i)}Sim(i,j)}.
\end{equation}

\subsubsection{\textbf{Hybrid Training/Prediction}} \quad\\
After the collaborative filtering prediction of the three-factor matrix, the factor prediction is achieved.
In this step, the predicted factors will be used to predict the unknown success rate.

\textbf{(1)Hybrid Training}

First, it is assumed that there is a mapping between Success Rate and the three factors (RightBlock, RecentHeight, and RoundTripTime):
\begin{equation}
\begin{split}
SuccessRate_{i,j} = f(&RightBlock_{i,j}, RecentHeight_{i,j},\\
&RoundTripTime_{i,j})
\end{split}
\end{equation}
Thus the mapping can be transferred to the matrix by the above equations.

And H-BRP sets up a regression model to fit this mapping. As shown in Figure \ref{brt}, as the gray area and arrows show, the known SuccessRate (gray area) and the known value in the three-factor matrices are used to train the regression model. During the training, the model learns from this hybrid data.

\textbf{(2)Success Rate Prediction}

After the regression training, the regression model can represent the mapping between Success Rate and the three factors (RightBlock, RecentHeight, and RoundTripTime).
Thus the factors predicted in the collaborative prediction can be input into the model, with the output as the success rate. As shown in Figure \ref{brt}, the predicted three-factors matrices (the white area) are input into the regression model and come out with the SuccessRate matrix predicted (the yellow area).

\textbf{(3)Predict Reliabity}

By the above steps, H-BRP obtain the predicted Success Rate from blockchain requester \emph{R$_i$} to Blockchain Peer \emph{P$_j$}. 
To predict the reliability of \emph{P$_j$} observed by \emph{R$_i$},
H-BRP adopts the commonly used exponential reliability function~\cite{lyu96handbook}:
\begin{equation}
Reliability_{i,j}(t) = e^{- \gamma \times t},
\end{equation} where $\gamma$ (\emph{failure-rate}) is the rate
of failures of request during a certain time duration, and $t$ is the time period for which the reliability is to be calculated.

The value of $\gamma$ can be calculated by:
\begin{equation}
\gamma = 1 - SuccessRate_{i,j}
\end{equation}
Thus the reliability from \emph{R$_i$} to \emph{P$_j$} can be calculated by:
\begin{equation}\label{rate2relia}
Reliability_{i,j}(t) = e^{-(1 - SuccessRate_{i,j}) \times t}.
\end{equation}

\section{Implement and Experiment} \label{Implement and Experiment}
In this section, we implement and evaluate the proposed approach based on a real-world dataset, which is collected from 100 requesters to 200 blockchain peers. First, we introduce the details of implementation and dataset description, and then the evaluation \& analysis of three research questions (i.e., reliability, accuracy, parameters impacts) are introduced, respectively.

\subsection{Implement and Dataset}
H-BRP is implemented by ShellScript, NodeJS and Python. 
Random Batch Block Request Testing is implemented by ShellScript to enable it to collect the data in all Linux server.
Although there are some Remote Procedure Call testing frameworks that can be used, but most of them have lots of dependencies. And the dependencies are different in different Linux versions case by case.
If a user wants to install H-BRP quickly to his client, the program should be light enough. ShellScript can meet all these requirements.
And Matrix Generation is implemented by NodeJS and Python. More specifically, the NodeJS program is used to parse and analysis the data from the main blockchain to check which block request returns the right block. And the Python program is used to generate the SuccessRate Matrix, RightBlock Matrix, RecentHeight Matrix, and RoundTripTime Matrix and predict the blockchain reliability. 

As for the blockchain requesters and peers.
PlanetLab\footnote{http://www.planetlab.org} is an organization that provided more than 1000 nodes all over the world. In this paper, 61 of them are selected to send the block requests.
Vultr\footnote{http://www.vultr.com} is a platform that provides cloud server leases. In this paper, 35 Linux servers (running the Cent OS) are rented from Vultr.
Besides another 4 Linux servers owned by the research team, H-BRP deploys the requester program on 100 servers in total as the requesters.
Ethernode\footnote{http://www.ethernode.org} is a website that showing all the blockchain peers of Ethereum over the world. In this paper, 200 blockchain peers are selected to be test. The blockchain peers are from 21 countries and the requesters are from 15 countries.

We deploy the requester with the batch size as \emph{n=5} and the time period as \emph{t=5}. After deploying the requesters, each requester sends random batch block requests to 5 blockchain peers in the period of 5 seconds. Finally, with the H-BRP implement, a dataset of over 2,000,000 test cases from 100 requesters to 200 blockchain peers is obtained. All the implement and dataset will be released on the website. For double-blind review, we upload the examination result to an anonymous github\footnote{https://github.com/forreview/H-BRP \label{thegithub}}.

The experiment of analysis and prediction is conducted on the dataset to answer the following research questions:

\textbf{Question 1: }How is the reliability of the blockchain system evaluated by H-BRP?

\textbf{Question 2: }How accurate is the method proposed compared with other reliability prediction methods?

\textbf{Question 3: }What is the impact of different parameters set in the model?

\subsection{RQ1: Case Study} \label{rq1}
In this subsection, H-BRP parses and analysis the obtained dataset to give some cases study to see the reliability of the blockchain system. 
The matrix generation is under the experimental settings of \emph{MaxBlockBack=12}, \emph{MaxRTT=2000}. 
After matrix generation, the dataset is presented as a 100 $\times$ 200 SuccessRate matrix. 

\begin{figure}
\centering
\includegraphics[width=3.4in]{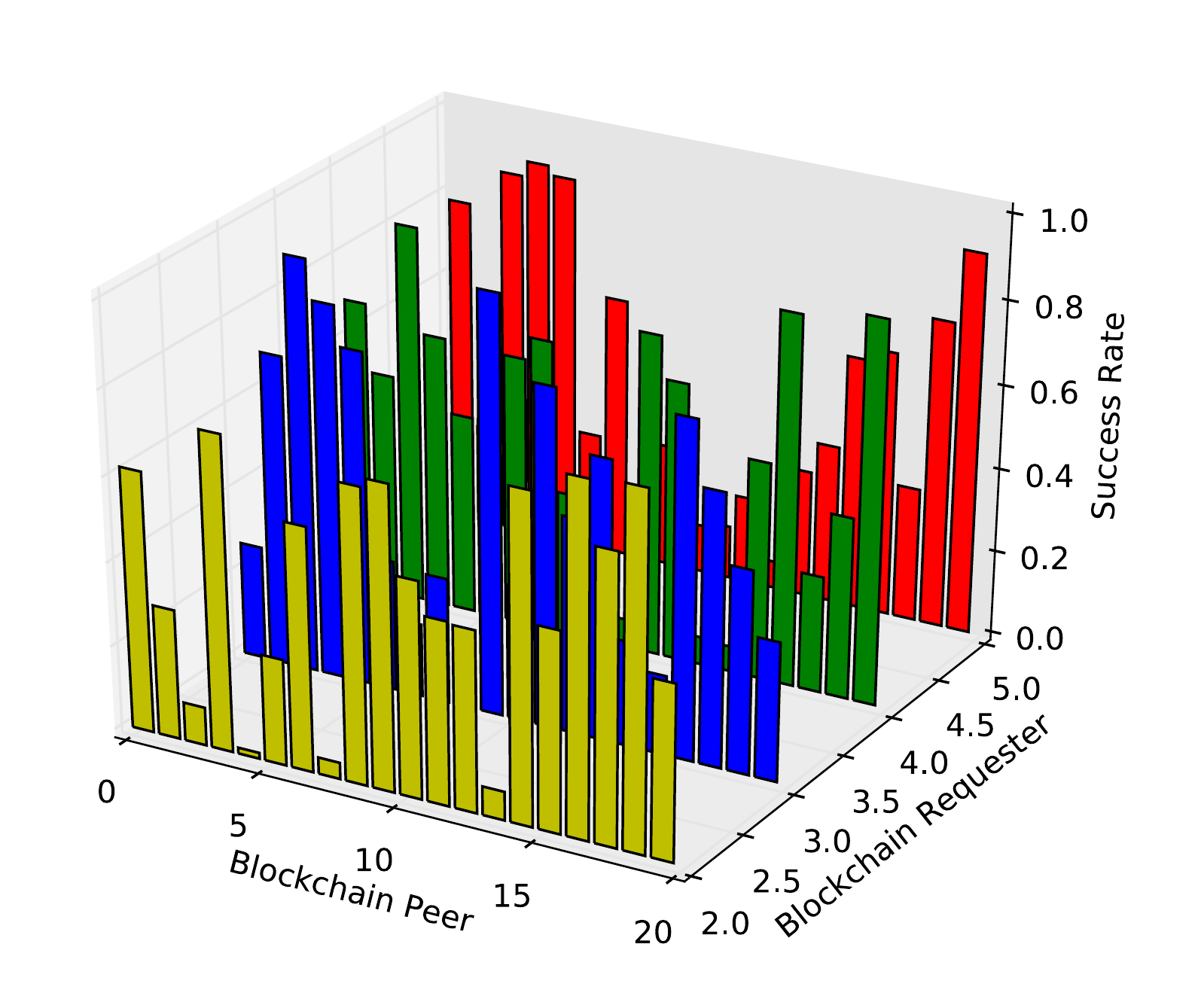}
\caption{Success Rate Distribution of H-BRP Dataset} \label{casestudy}
\end{figure}

From Equation \ref{rate2relia} we learn that the higher success rate means the higher reliability.
Figure~\ref{casestudy} shows the success rate distribution of 20 blockchain peers and 4 requesters.
In this case, the blockchain peers show different reliability to different requesters.
This is mainly caused by the network situation that some requesters cannot receive the block from some remote peers due to the long network delay
Thus the reliability they observed could be different.
Since different requesters have different observed reliability to the same peer, it is required to make personalized prediction.

\begin{table}
\caption{Case Study of H-BRP Dataset}\label{case}
\centering
\begin{tabular}{|l|ccc|}
\hline
\diagbox{Requester}{Success Rate}{Peer} & {\bf 147.75.111.247} & {\bf 147.75.100.193} & ...\\ \hline
{\bf 130.194.252.8} & 0.4873 & 0.0802 & ...\\
{\bf 130.194.252.9} & 0.4444 & 0.0327& ...\\
{\bf 192.33.90.67} & 0.1783 & 0.7079 & ...\\
{\bf 194.29.178.14} & 0.1929 & 0.7014 & ...\\ 
...& ...& ...& ... \\\hline
\end{tabular}
\end{table}

\begin{figure}
\centering
\includegraphics[width=3.2in]{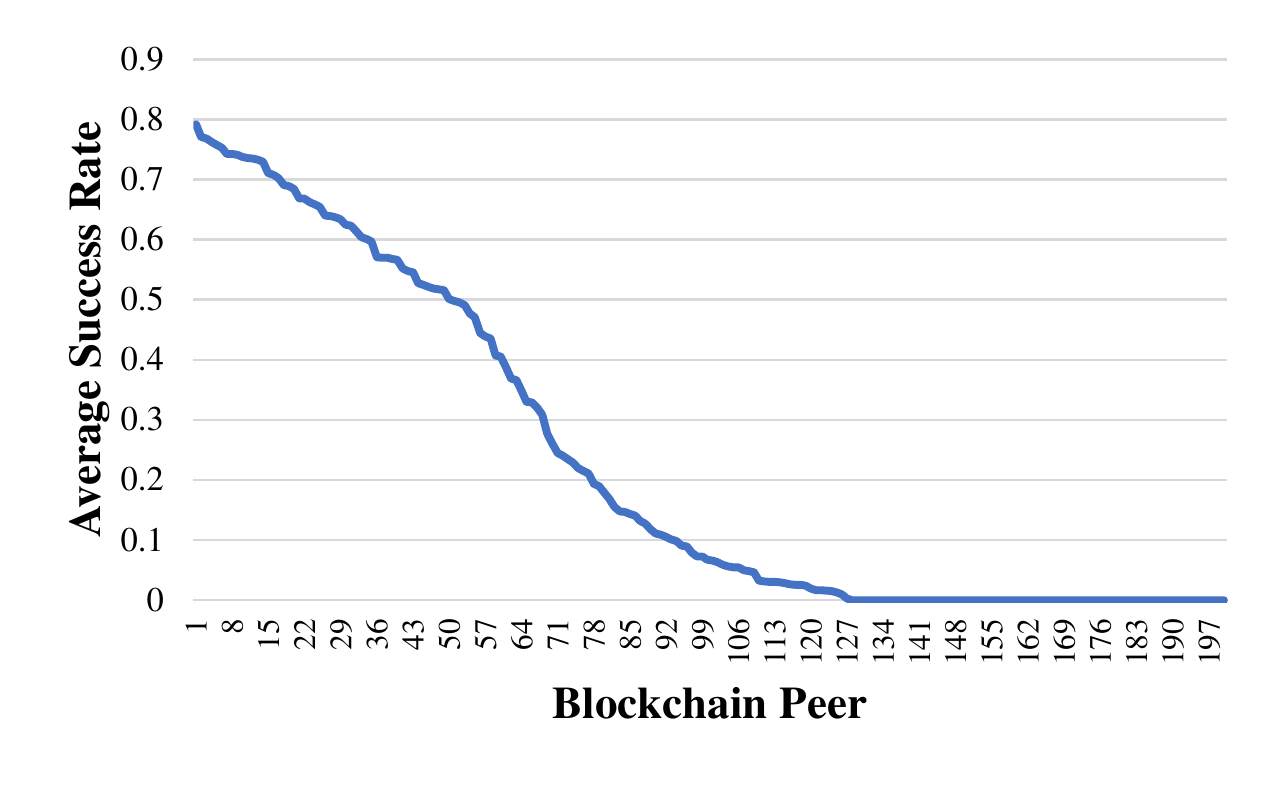}
\caption{Average Succsess Rate of Blockchain Peers} \label{peerrelia}
\end{figure}

\begin{figure}
\centering
\includegraphics[width=3.2in]{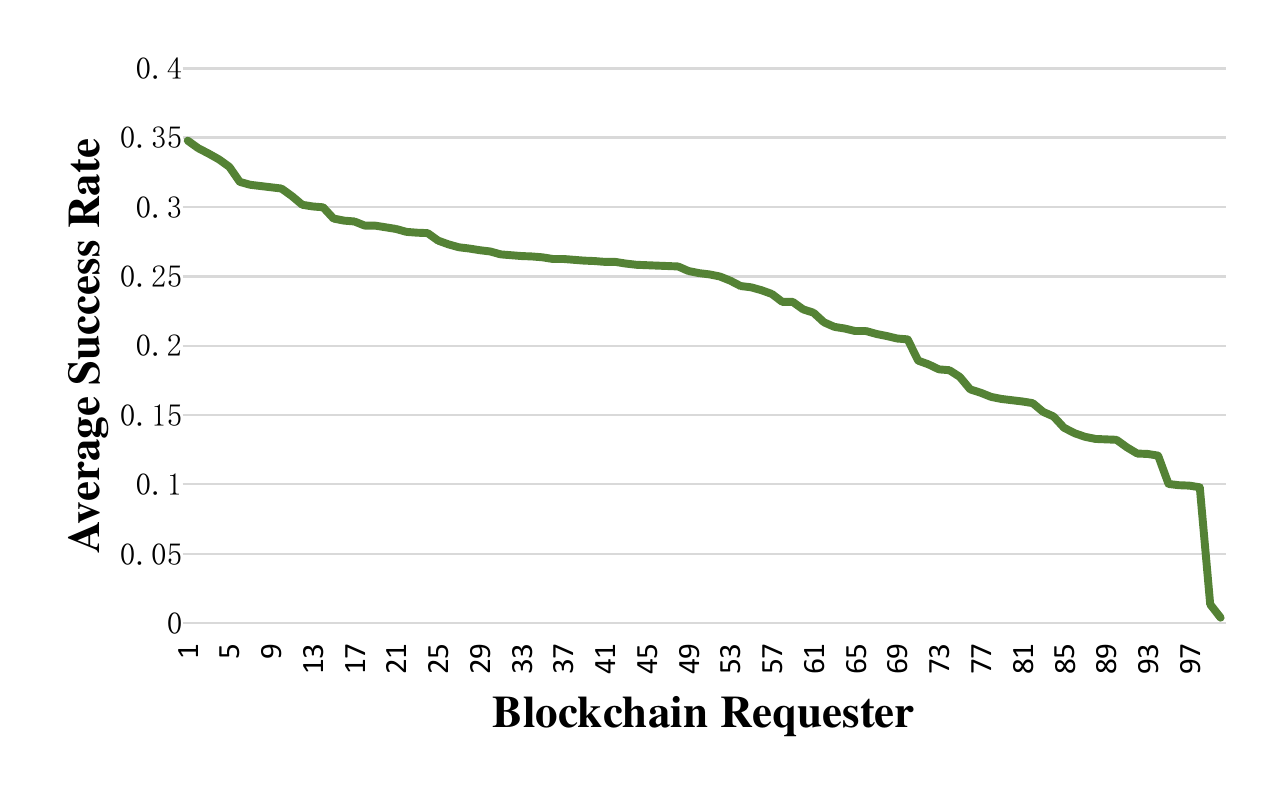}
\caption{Average Succsess Rate of Blockchain Requesters} \label{userrelia}
\end{figure}

Moreover, we extract 2 peers and 4 requesters to see how is the reliability exactly, as shown in Table \ref{case}.
As for the requesters in 130.194.252.8 and 130.194.252.9, the success rate when they connect to 147.75.111.247 is much higher than 147.75.100.193.
However, when considering to 192.33.90.67 and 194.29.178.14, the comparison of success rate is opposite.
From this case, we can learn that the similar users may observe similar reliability of similar peers.
That is why H-BRP obtains the relationship between similar users and peers to do collaborative prediction.

\begin{table*}
\caption{Comparison of RMSE of Blockchain Reliability Prediction Approaches}\label{q2}
\centering
\begin{tabular}{|c|c|c|c|c|c|c|}
\hline
\bf{Parameter} & \bf{Method} & \bf{Density=0.30} & \bf{Density=0.50} & \bf{Density=0.65} & \bf{Density=0.80} & \bf{Density=0.95} \\ \hline
&\bf{UMEAN} & 0.3789 & 0.3774 & 0.3765 & 0.3758 & 0.3755 \\
&\bf{IMEAN} & 0.0925 & 0.0922 & 0.0925 & 0.0921 & 0.0918 \\
\bf{MaxBlockBack=0,} &\bf{UPCC} & 0.2823 & 0.2791 & 0.2769 & 0.2754 & 0.2758 \\
\bf{MaxRTT=1000}&\bf{IPCC} & 0.0821 & 0.0777 & 0.0764 & 0.0758 & 0.0748 \\
&\bf{UIPCC} & 0.0851 & 0.0806 & 0.0791 & 0.0782 & 0.0773 \\
&\bf{H-BRP} & \bf{0.0803} & \bf{0.0731} & \bf{0.0712} & \bf{0.07} & \bf{0.0672} \\ \hline

&\bf{UMEAN} & 0.4547 & 0.4531 & 0.4519 & 0.4513 & 0.4508\\
&\bf{IMEAN} & 0.1171 & 0.1168 & 0.1167 & 0.1162 & 0.1166\\
\bf{MaxBlockBack=12,}&\bf{UPCC} & 0.3627 & 0.3591 & 0.3566 & 0.3552 & 0.3559\\
\bf{MaxRTT=1000}&\bf{IPCC} & \bf{0.1009} & 0.0949 & 0.0919 & 0.0908 & 0.092\\
&\bf{UIPCC} & 0.1053 & 0.0994 & 0.0963 & 0.0951 & 0.0961\\
&\bf{H-BRP} & 0.1031 & \bf{0.0925} & \bf{0.0899} & \bf{0.0879} & \bf{0.0845}\\ \hline

&\bf{UMEAN} & 0.4735 & 0.472 & 0.4707 & 0.4702 & 0.47\\
&\bf{IMEAN} & 0.0848 & 0.0851 & 0.0848 & 0.0845 & 0.085\\
\bf{MaxBlockBack=12,}&\bf{UPCC} & 0.3793 & 0.3775 & 0.3758 & 0.3752 & 0.376\\
\bf{MaxRTT=2000}&\bf{IPCC} & 0.081 & 0.0785 & 0.0765 & 0.0748 & 0.0751\\
&\bf{UIPCC} & 0.0887 & 0.0864 & 0.0845 & 0.0829 & 0.0831\\
&\bf{H-BRP} & \bf{0.0654} & \bf{0.0567} & \bf{0.0529} & \bf{0.0507} & \bf{0.0464}\\ \hline

&\bf{UMEAN} & 0.5088 & 0.5081 & 0.5067 & 0.5065 & 0.5067\\
&\bf{IMEAN} & 0.0938 & 0.094 & 0.0934 & 0.0931 & 0.0941\\
\bf{MaxBlockBack=100,}&\bf{UPCC} & 0.4595 & 0.4585 & 0.4569 & 0.4564 & 0.4574\\
\bf{MaxRTT=5000}&\bf{IPCC} & 0.0921 & 0.0887 & 0.0842 & 0.0806 & 0.0774\\
&\bf{UIPCC} & 0.1022 & 0.0993 & 0.0954 & 0.0923 & 0.0895\\
&\bf{H-BRP} & \bf{0.0648} & \bf{0.0562} & \bf{0.0524} & \bf{0.0494} & \bf{0.0433}\\ \hline
\end{tabular}
\end{table*}

We rank the blockchain peers to see how their reliability is. The average success rate of block requests is shown in Figure~\ref{peerrelia}.
In these 200 blockchain peers, half of them show very low reliability to all the requesters, which means that they do not always return the latest block in time. This is mainly caused by that the block propagation in blockchain system is slow. Once a block is mined, it takes it a period of time to be propagated to the whole network. If the P2P network connectivity is not good, some of the peers will not receive the latest block.
On the other hand, there are some large blockchain miners that generate most of the blocks. Thus the peers that are closer to the miners will have the higher chance to receive the latest block, resulted in the difference of reliability.

As for the blockchain users, we calculate out the average success rate of 100 requesters to see the reliability observed by the users.
Figure~\ref{userrelia} shows the result that the average success rate is lower than 0.3.
It means that, if a user connect to the blockchain peers randomly, his chance to get the correct latest block is quite low.
Compared to the most reliable peer shown in Figure~\ref{peerrelia}, if the user connect to the most reliable one, the chance to get the block will be increased by more than two times. Therefore, before selecting blockchain peers, predicting the reliability and choosing the reliable peers will truly help the users to get the latest block.

\subsection{RQ2: Accuracy of Different Method}

To study the prediction performance, we compare our approach (H-BRP) with five other ones in reliability prediction: user-mean (UMEAN), item-mean (IMEAN), user-based approach using PCC (UPCC)~\cite{breese:empirical}, item-based approach using PCC (IPCC)~\cite{sarwar:item-based}, and user-item-based approach (UIPCC)~\cite{zheng2010collaborative}. UMEAN employs the average success rate of the current requester on other blockchain peers for the prediction, while IMEAN employs the average success rate of the blockchain peers observed by other requesters for the prediction. UPCC only employs similar blockchain requesters for the failure probability prediction, while IPCC only employs similar blockchain peers for the prediction. And UIPCC is the combination of UPCC and IPCC. 
In this paper, those approaches are compared with H-BRP, to predict the same training Success Rate Matrix.

For each round, first we randomly remove the entries in the generated Success Rate Matrix to transfer it into the target density. After that, the removed entries are set as the test value.
The same training matrix is the input of every reliability prediction approach, while the predicted value is the output. And the output value is compared with the test value to measure the prediction accuracy.

Root Mean Square Error (RMSE) metric is employed to measure the prediction accuracy of different approaches. RMSE is defined as:
\begin{equation}
RMSE = \sqrt{\frac{\sum_{r,p} (SuccessRate_{r,p} - \widehat{SuccessRate}_{r,p})^2}{N}},
\end{equation}
where smaller RMSE values indicate better prediction accuracy.

\renewcommand\subfigcapskip{-0.5ex}
\begin{figure*}[htb]
\centering \subfigure[MaxBlock'=0, MaxRTT=1000]{
\centering
\includegraphics[width=4.1cm]{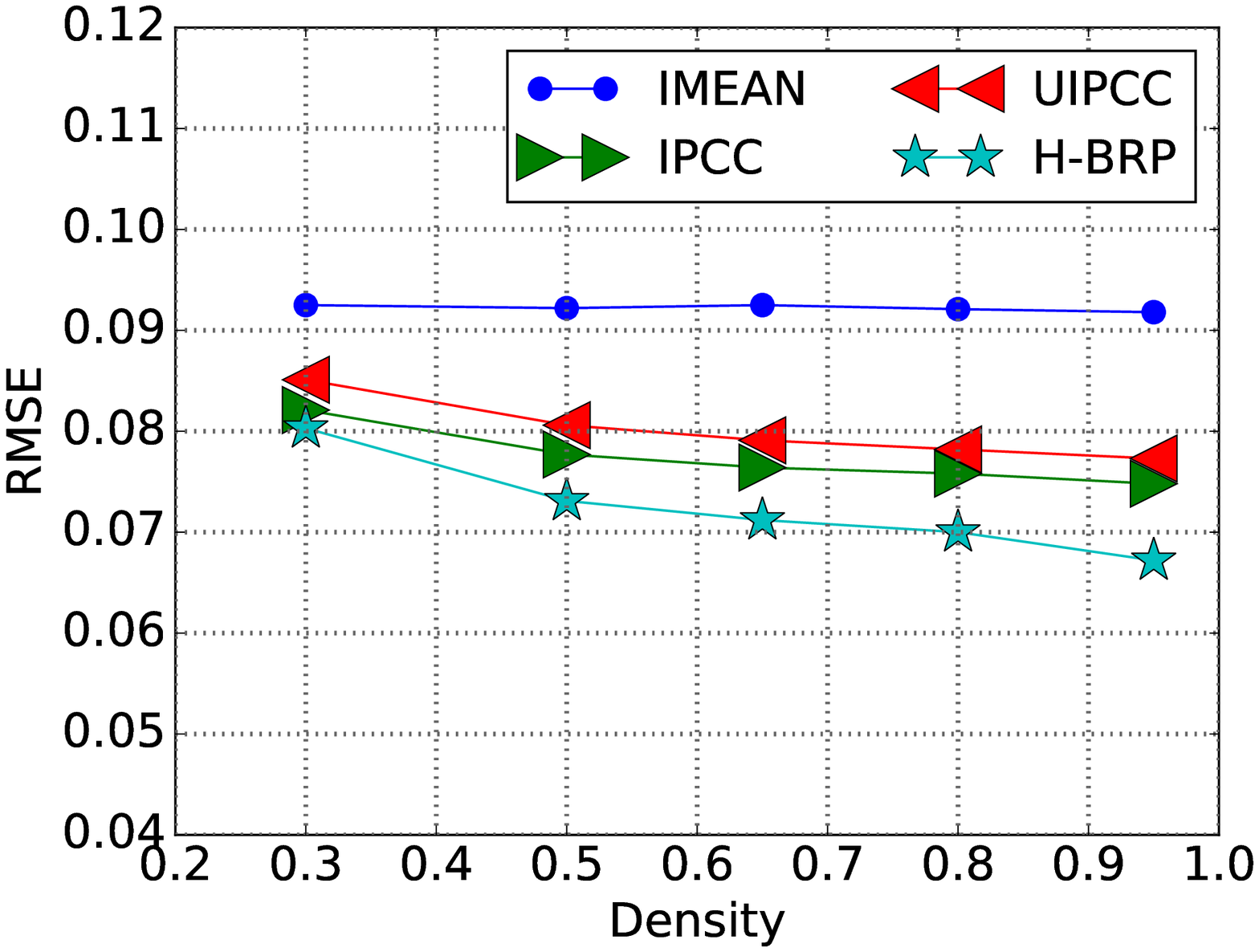}
} \subfigure[MaxBlock'=12, MaxRTT=1000]{
\centering
\includegraphics[width=4.1cm]{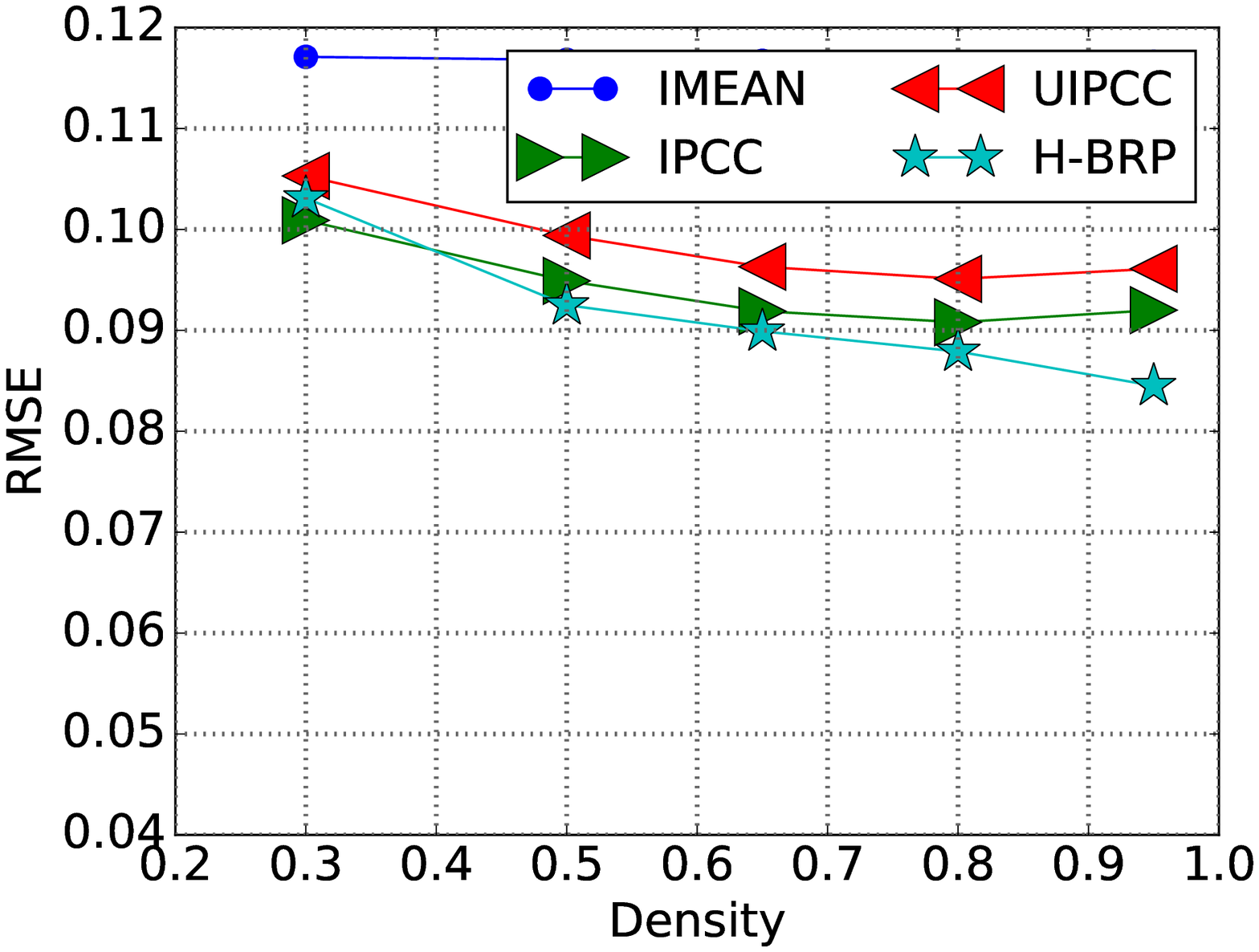}
} \subfigure[MaxBlock'=12, MaxRTT=2000]{
\centering
\includegraphics[width=4.1cm]{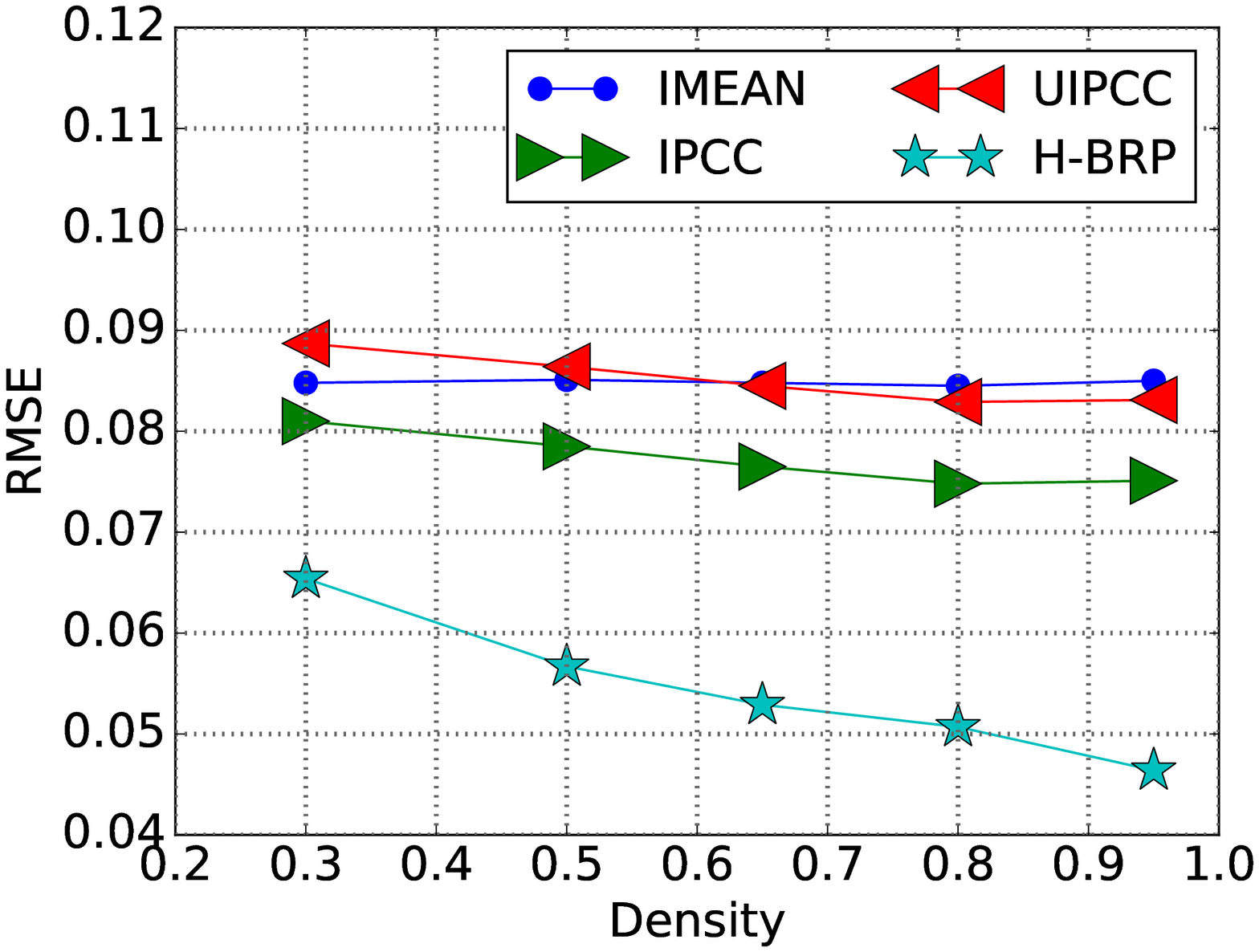}
} \subfigure[MaxBlock'=100, MaxRTT=5000]{
\centering
\includegraphics[width=4.1cm]{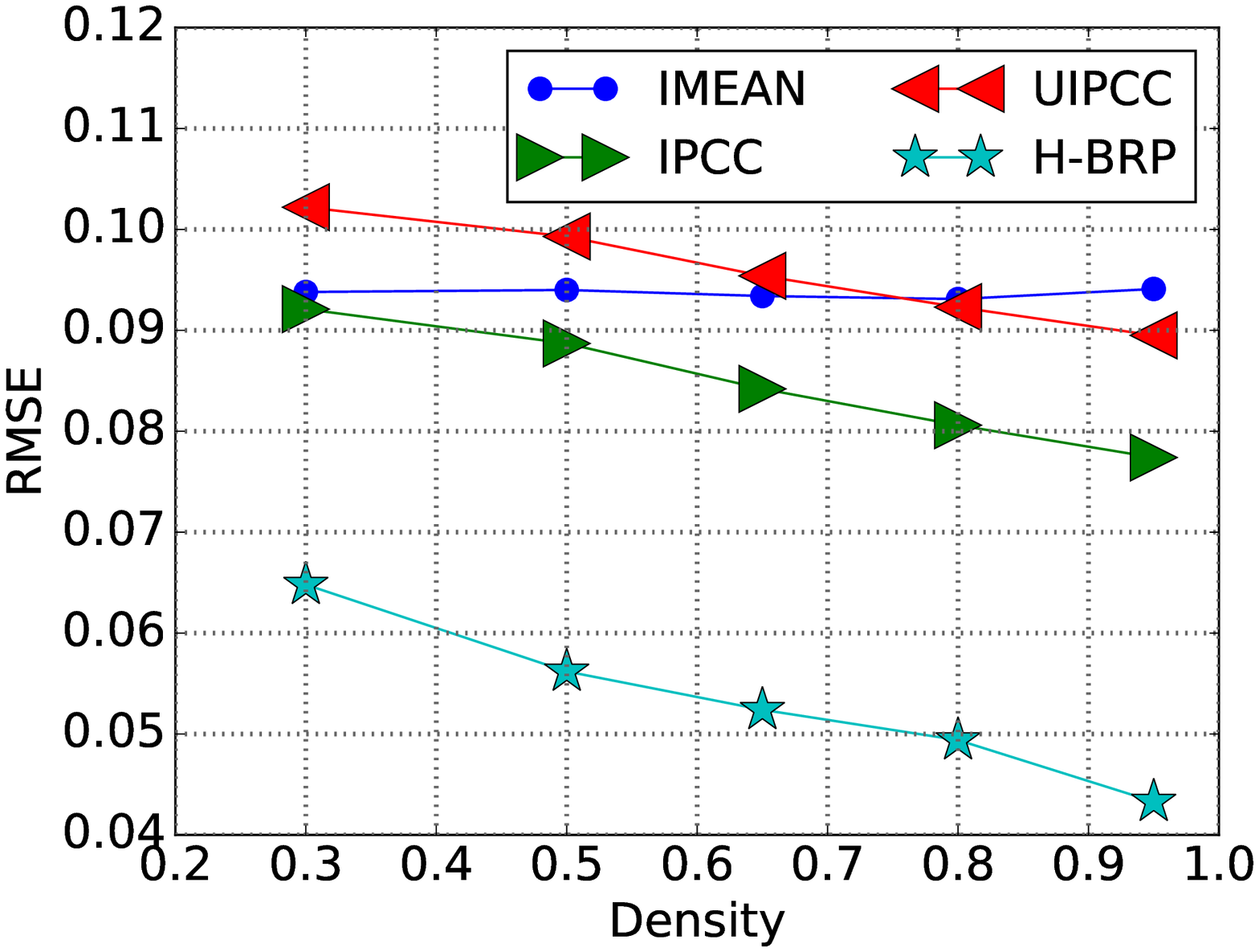}
} \caption{Impact of Density} \label{impa}
\end{figure*}

\renewcommand\subfigcapskip{-0.5ex}
\begin{figure}
\centering 
\subfigure[MaxRTT=2000, Density=0.5]{
\centering
\includegraphics[width=4cm]{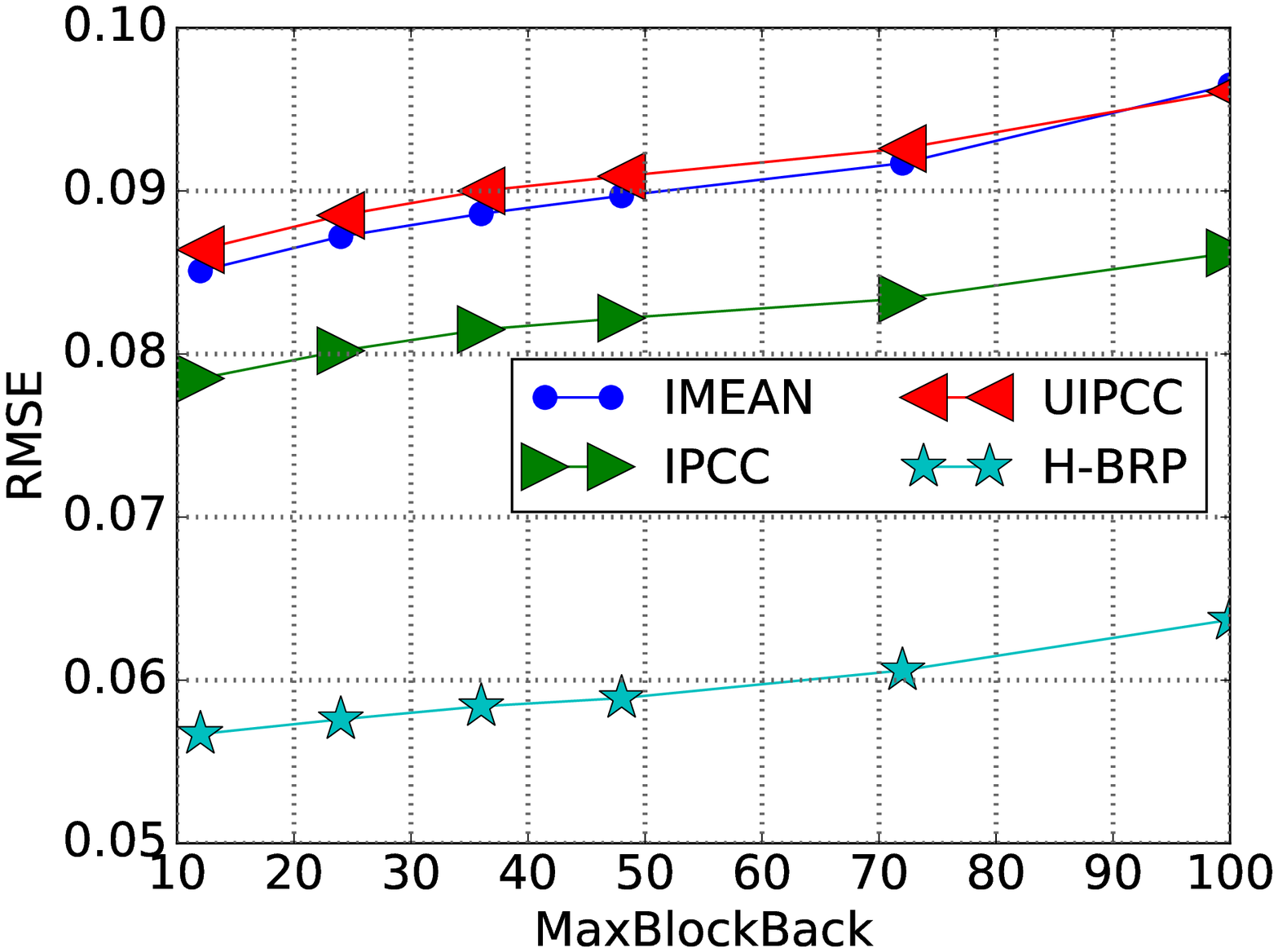}
} 
\subfigure[MaxRTT=2000, Density=0.8]{
\centering
\includegraphics[width=4cm]{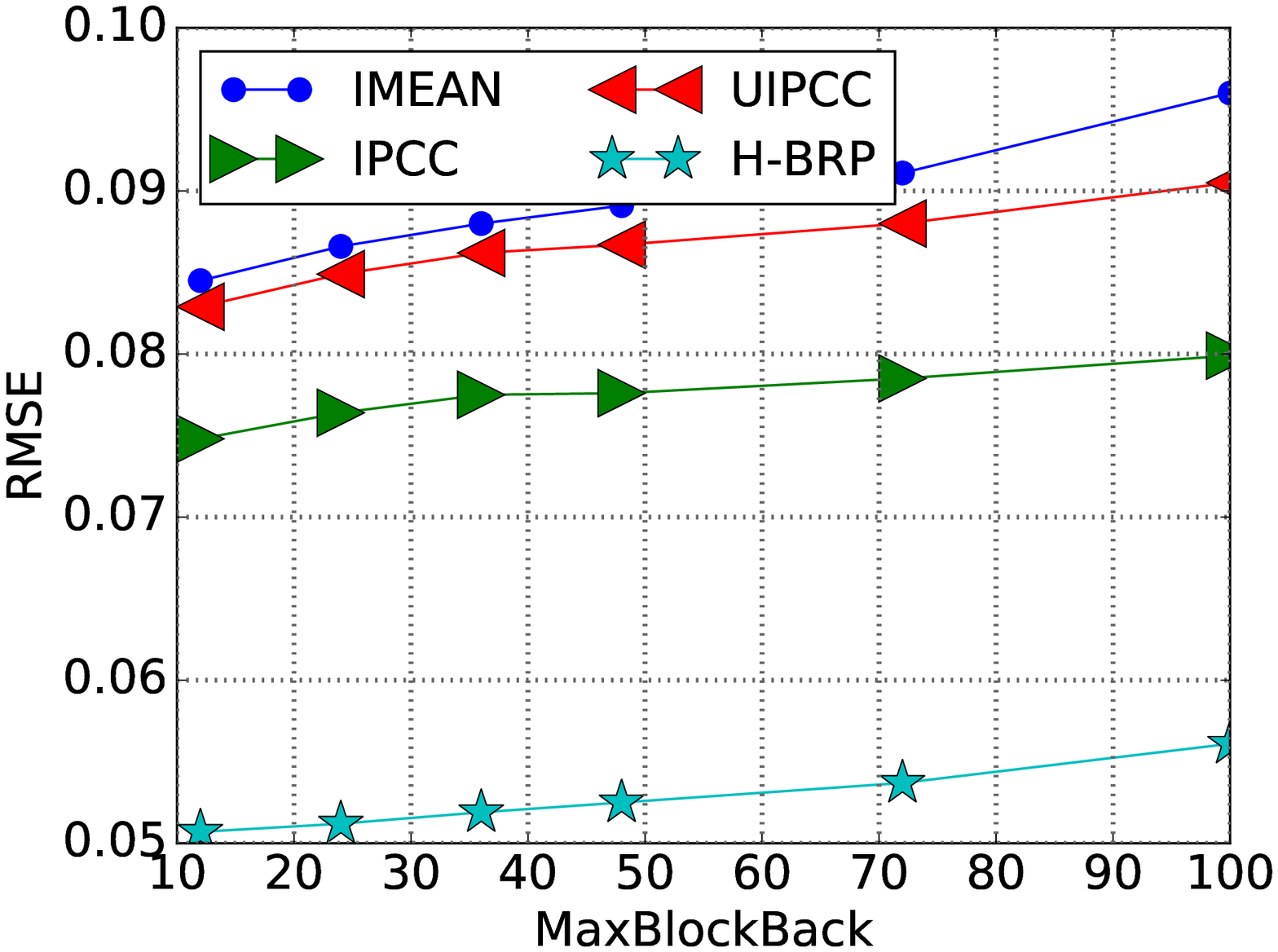}
} \caption{Impact of MaxBlockBack} \label{impablockback}
\end{figure}

\renewcommand\subfigcapskip{-0.5ex}
\begin{figure}
\centering 
\subfigure[MaxBlockBack=12, Density=0.5]{
\centering
\includegraphics[width=4cm]{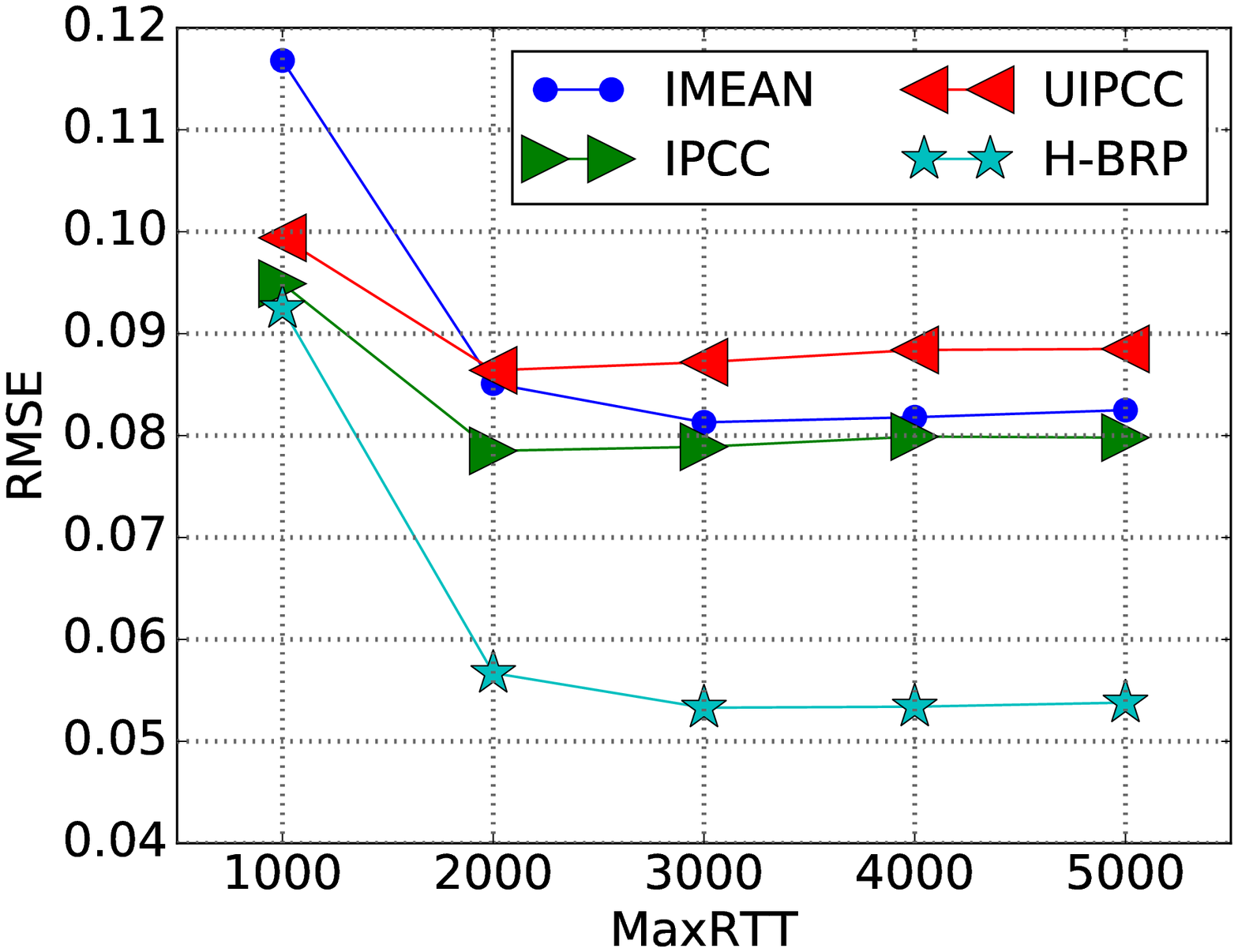}
} 
\subfigure[MaxBlockBack=12, Density=0.8]{
\centering
\includegraphics[width=4cm]{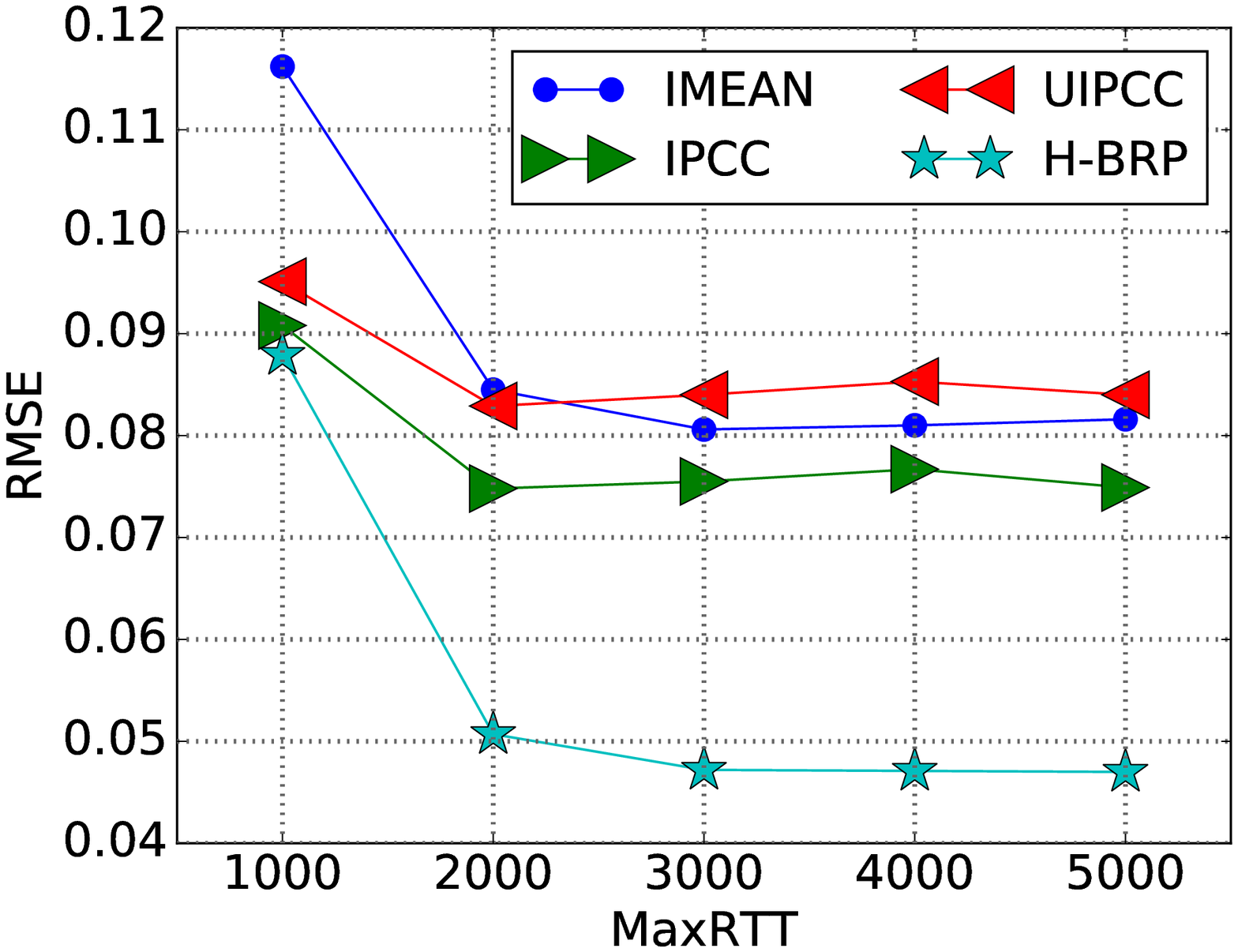}
} \caption{Impact of MaxRTT} \label{impartt}
\end{figure}

As for the parameters in this subsection, the density of the matrices is set as \emph{density = 0.3, 0.5, 0.65, 0.80, 0.95}.
We set \emph{K=3} to select Top3 similar blockchain peers for the collaborative prediction.
And we set \emph{$<$MaxBlockBack=0, MaxRTT=1000$>$} to evaluate the prediction accuracy for the siutation that have extremely high requirements for blockchain synchronization speed (e.g., Bitcoin miner).
We set \emph{$<$MaxBlockBack=12, MaxRTT=1000$>$} and \emph{$<$MaxBlockBack=12, MaxRTT=1000$>$} to evaluate the accuracy for the situation that have high requirement for confirming blockchain data (e.g., cryptocurrencies wallet, cryptocurrencies exchange).
We set \emph{$<$MaxBlockBack=100, MaxRTT=5000$>$} to evaluate the accuracy for daily usage (e.g., ordinary blockchain users) that has high tolerance for block backwardness and latency.

The experiment under the same setting will be run in 20 rounds then come out with the average value of RMSE. The results are shown in Table \ref{q2}.
The experiment results show that H-BRP model achieves better accuracy than other approaches in different requirements for reliability and different matrix density. Mean Absolute Error (MAE) and Normalized Mean Absolute Error (NMAE) are also used in this experiment. The results can be checked on the anonymous github\textsuperscript{\ref{thegithub}}.

\subsection{RQ3: Impact of Parameters}

In this subsection, comparison of RMSE with different parameters is given to evaluate the impact of different parameters set in the model.
Since UPCC and UMEAN have much higher RMSE (lower accuracy) than other approaches, we will not take UPCC and UMEAN into the comparison to make the figures more clear.

\subsubsection{\textbf{Impact of Density}} \quad\\

In this experiment, we compare the RMSE in the same \emph{MaxBlockBack} and \emph{MaxRTT} to see the impact of the density of the training matrix.

As shown in Figure \ref{impa}, the results show that the accuracy rises as density increases. The main reason is that, the higher density of the training matrix is, the more information is input into the model, thus the more accurate the model is.
The result also shows that H-BRP model has better accuracy than other models in most cases. It means that even each requester only has the request history with random 30\% of blockchain peers, the prediction of the remain 70\% can be realized.

\subsubsection{\textbf{Impact of MaxBlockBack}} \quad\\

\emph{MaxBlockBack} represents the block backwardness tolerance of the blockchain requesters.
To compare the impact of \emph{MaxBlockBack}, we set the parameters as \emph{MaxRTT=2000} and \emph{Density=0.5, 0.8} to see how the accuracy is changed with the variance of block backwardness tolerance.

The experiment result shows that the more block backwardness tolerance given, the lower accurate the model is, as the RMSE is increasing. 
This may be caused by that the higher block latency tolerance is given, the less difference between the blockchain peers is. Thus the accuracy of the models is affected.

\subsubsection{\textbf{Impact of MaxRTT}} \quad\\

As for different \emph{MaxRTT}, we set the experimental parameters as \emph{MaxBlockBack=12} and \emph{Density=0.5, 0.8} to see how the accuracy is changed with the round-trip time tolerance.

The experiment result shows that the higher \emph{MaxRTT} is given, the more accurate the model is. 
In the \emph{MaxRTT$ = $1000}, the RMSE of all prediction models are very close and large, which means that the models show low accuracy in this setting. The main reason is that only few blockchain peers can response in 1000 ms. Therefore, the fluctuation of success rate is relatively large, which causes the models to be less accurate.
However, especially in the situation that \emph{MaxRTT$ > $1000}, H-BRP shows great advantage over other models.

\section{Related Work and Discussion} \label{Related Work and Discussion}
This section will describe the related work in blockchain reliability prediction, including blockchain related research and traditional software reliability research.

As for the blockchain reliability or availability, Zheng et.al \cite{zheng2018detailed} propose a scalable framework for detailed and real-time monitoring of blockchain systems, which has much lower overhead and more details about the blockchain systems compared with previous approaches. 
Weber et al. \cite{weber2017availability} propose a method to identify the availability limitations of Bitcoin and Ethereum, showing that the reading availability is high while the writing availability is low. Kalodner et al. \cite{kalodner2017blocksci} propose an open-source software platform for blockchain systems, which parsing the data from the p2p nodes and raw blockchain data for users to monitor and analyze the system. Yang et al. \cite{blockguide} propose a benchmark for Fabric blockchain. Dinh et al. \cite{dinh2017blockbench} describe frameworks for analyzing private blockchains in varying workloads. Guapta et al. \cite{analyzingperform} also propose a method for analyzing performance. Gervais et al. \cite{gervais2016security} present a novel quantitative framework for the security and performance of PoW blockchains.

As for traditional software reliability research, Michael et al. propose a handbook of software reliability engineering \cite{lyu96handbook}. The main idea of software reliability prediction is to predict the unknown reliability of software systems based on the past data \cite{DBLP:conf/sigsoft/JiangZLSHGS13}. Chen et al. \cite{chen2011enhanced} propose an enhanced qos prediction approach for service selection.
Zheng et al. \cite{zheng2010collaborative} and Silic et al. \cite{DBLP:conf/sigsoft/SilicDS13} propose a set of collaborative filtering approaches to predict reliability of software systems.

However, the previous blockchain research does not give a method of reliability prediction for blockchain systems. And it always focus on few blockchain peers. On the other hand, the previous research about reliability prediction cannot fit the blockchain systems since blockchain factors are not taken into consideration. To attack these challenges, in this paper, the main idea of Hybrid Blockchain Reliability Prediction model is to extract blockchain related factors to predict the reliability of blockchain system.

\section{Conclusion and Future Work} \label{Conclusion and Future Work}

In this paper, we firstly propose a Hybrid Blockchain Reliability Prediction model for blockchain systems. It can do personalized reliability prediction for blockchain users to improve the block synchronization speed and avoid the loss of cryptocurrencies. Real-world experiment with 2,000,000 test cases from 100 requesters to 200 blockchain peers is conducted, and the results show the proposed model is effective and more accurate than previous reliability prediction model.
Specifically, implementation details and the dataset will be released for research.

In the future, our work can be extended in different aspects:
\textbf{(1)~Decentralized Collector: } The centralized collector is a limitation as a blockchain tool. To collect the data on blockchain could be chosen but the throughput would be too low. It should be transferred to a suitable decentralized platform. 
\textbf{(2)~Model Complexity: }As for time complexity, H-BRP consumes 120\% to 530\% of other approaches in different scalability. This might not meet the requirement of online prediction. The model complexity could be improved.
\textbf{(3)~Scalability: }This paper only selects 200 blockchain peers of Ethereum Mainnet to evaluate and predict the reliability. However, it is reported that there are over 14,000 Ethereum peers and over 10,000 Bitcoin peers over the world which are available to be test.


\bibliographystyle{ACM-Reference-Format}
\bibliography{yinyong}


\begin{thebibliography}{24}


\ifx \showCODEN    \undefined \def \showCODEN     #1{\unskip}     \fi
\ifx \showDOI      \undefined \def \showDOI       #1{#1}\fi
\ifx \showISBNx    \undefined \def \showISBNx     #1{\unskip}     \fi
\ifx \showISBNxiii \undefined \def \showISBNxiii  #1{\unskip}     \fi
\ifx \showISSN     \undefined \def \showISSN      #1{\unskip}     \fi
\ifx \showLCCN     \undefined \def \showLCCN      #1{\unskip}     \fi
\ifx \shownote     \undefined \def \shownote      #1{#1}          \fi
\ifx \showarticletitle \undefined \def \showarticletitle #1{#1}   \fi
\ifx \showURL      \undefined \def \showURL       {\relax}        \fi
\providecommand\bibfield[2]{#2}
\providecommand\bibinfo[2]{#2}
\providecommand\natexlab[1]{#1}
\providecommand\showeprint[2][]{arXiv:#2}

\bibitem[\protect\citeauthoryear{??}{imt}{[n. d.]}]%
        {imtokenlate}
 \bibinfo{year}{[n. d.]}\natexlab{}.
\newblock \bibinfo{booktitle}{\emph{Own Your RPC: Abnormal Wallet Peer Causes
  Money Lost by Repeated Transaction}}.
\newblock
  https://medium.com/c2736464697/own-your-rpc-abnormal-wallet-peer-causes-money-lost-by-repeated-transaction-9aa7fefea8e6.
\newblock


\bibitem[\protect\citeauthoryear{??}{wha}{2017}]%
        {what3}
 \bibinfo{year}{2017}\natexlab{}.
\newblock \bibinfo{booktitle}{\emph{What is a Decentralized Application
  CoinDesk}}.
\newblock
  https://www.\\coindesk.com/information/what-is-a-decentralized-application-dapp/.
\newblock


\bibitem[\protect\citeauthoryear{??}{You}{2017}]%
        {YourfirstDapp}
 \bibinfo{year}{2017}\natexlab{}.
\newblock \bibinfo{booktitle}{\emph{Your first Dapp}}.
\newblock https://dappsforbeginners.wordpress.com/tutorials/your-first-dapp/.
\newblock


\bibitem[\protect\citeauthoryear{Breese, Heckerman, and Kadie}{Breese
  et~al\mbox{.}}{1998}]%
        {breese:empirical}
\bibfield{author}{\bibinfo{person}{John~S. Breese}, \bibinfo{person}{David
  Heckerman}, {and} \bibinfo{person}{Carl Kadie}.}
  \bibinfo{year}{1998}\natexlab{}.
\newblock \showarticletitle{Empirical Analysis of Predictive Algorithms for
  Collaborative Filtering}. In \bibinfo{booktitle}{\emph{Proc. 14th Annual
  Conf. Uncertainty in Artificial Intelligence (UAI'98)}}.
  \bibinfo{pages}{43--52}.
\newblock


\bibitem[\protect\citeauthoryear{Buterin et~al\mbox{.}}{Buterin
  et~al\mbox{.}}{2013}]%
        {buterin2013ethereum}
\bibfield{author}{\bibinfo{person}{Vitalik Buterin} {et~al\mbox{.}}}
  \bibinfo{year}{2013}\natexlab{}.
\newblock \bibinfo{title}{Ethereum white paper}.
\newblock
\newblock


\bibitem[\protect\citeauthoryear{Chen, Feng, Wu, and Zheng}{Chen
  et~al\mbox{.}}{2011}]%
        {chen2011enhanced}
\bibfield{author}{\bibinfo{person}{Liang Chen}, \bibinfo{person}{Yipeng Feng},
  \bibinfo{person}{Jian Wu}, {and} \bibinfo{person}{Zibin Zheng}.}
  \bibinfo{year}{2011}\natexlab{}.
\newblock \showarticletitle{An enhanced qos prediction approach for service
  selection}. In \bibinfo{booktitle}{\emph{Services Computing (SCC), 2011 IEEE
  International Conference on}}. IEEE, \bibinfo{pages}{727--728}.
\newblock


\bibitem[\protect\citeauthoryear{Chen.}{Chen.}{2017}]%
        {Stranford}
\bibfield{author}{\bibinfo{person}{Richard Chen.}}
  \bibinfo{year}{2017}\natexlab{}.
\newblock \bibinfo{booktitle}{\emph{A Brief Overview of dApp Development}}.
\newblock
  https://thecontrol.co/a-brief-overview-of-dapp-development-b8ac1648322c.
\newblock


\bibitem[\protect\citeauthoryear{Dickson}{Dickson}{[n. d.]}]%
        {analyzingperform}
\bibfield{author}{\bibinfo{person}{Anuj Das Gupta.~Andrew Dickson}.}
  \bibinfo{year}{[n. d.]}\natexlab{}.
\newblock \bibinfo{title}{Analyzing Performance in Blockchain-Based Systems}.
\newblock
\newblock


\bibitem[\protect\citeauthoryear{Dinh, Wang, Chen, Liu, Ooi, and Tan}{Dinh
  et~al\mbox{.}}{2017}]%
        {dinh2017blockbench}
\bibfield{author}{\bibinfo{person}{Tien Tuan~Anh Dinh}, \bibinfo{person}{Ji
  Wang}, \bibinfo{person}{Gang Chen}, \bibinfo{person}{Rui Liu},
  \bibinfo{person}{Beng~Chin Ooi}, {and} \bibinfo{person}{Kian-Lee Tan}.}
  \bibinfo{year}{2017}\natexlab{}.
\newblock \showarticletitle{BLOCKBENCH: A Framework for Analyzing Private
  Blockchains}. In \bibinfo{booktitle}{\emph{Proceedings of the 2017 ACM
  International Conference on Management of Data}}. ACM,
  \bibinfo{pages}{1085--1100}.
\newblock


\bibitem[\protect\citeauthoryear{Gervais, Karame, W{\"u}st, Glykantzis,
  Ritzdorf, and Capkun}{Gervais et~al\mbox{.}}{2016}]%
        {gervais2016security}
\bibfield{author}{\bibinfo{person}{Arthur Gervais}, \bibinfo{person}{Ghassan~O
  Karame}, \bibinfo{person}{Karl W{\"u}st}, \bibinfo{person}{Vasileios
  Glykantzis}, \bibinfo{person}{Hubert Ritzdorf}, {and} \bibinfo{person}{Srdjan
  Capkun}.} \bibinfo{year}{2016}\natexlab{}.
\newblock \showarticletitle{On the security and performance of proof of work
  blockchains}. In \bibinfo{booktitle}{\emph{Proceedings of the 2016 ACM SIGSAC
  Conference on Computer and Communications Security}}. ACM,
  \bibinfo{pages}{3--16}.
\newblock


\bibitem[\protect\citeauthoryear{Jiang, Zhang, Liu, Song, Hung, Gu, and
  Sun}{Jiang et~al\mbox{.}}{2013}]%
        {DBLP:conf/sigsoft/JiangZLSHGS13}
\bibfield{author}{\bibinfo{person}{Yu Jiang}, \bibinfo{person}{Hehua Zhang},
  \bibinfo{person}{Han Liu}, \bibinfo{person}{Xiaoyu Song},
  \bibinfo{person}{William N.~N. Hung}, \bibinfo{person}{Ming Gu}, {and}
  \bibinfo{person}{Jiaguang Sun}.} \bibinfo{year}{2013}\natexlab{}.
\newblock \showarticletitle{System reliability calculation based on the
  run-time analysis of ladder program}. In \bibinfo{booktitle}{\emph{Joint
  Meeting of the European Software Engineering Conference and the {ACM}
  {SIGSOFT} Symposium on the Foundations of Software Engineering, ESEC/FSE'13,
  Saint Petersburg, Russian Federation, August 18-26, 2013}}.
  \bibinfo{pages}{695--698}.
\newblock
\urldef\tempurl%
\url{https://doi.org/10.1145/2491411.2494570}
\showDOI{\tempurl}


\bibitem[\protect\citeauthoryear{Kalodner, Goldfeder, Chator, M{\"o}ser, and
  Narayanan}{Kalodner et~al\mbox{.}}{2017}]%
        {kalodner2017blocksci}
\bibfield{author}{\bibinfo{person}{Harry Kalodner}, \bibinfo{person}{Steven
  Goldfeder}, \bibinfo{person}{Alishah Chator}, \bibinfo{person}{Malte
  M{\"o}ser}, {and} \bibinfo{person}{Arvind Narayanan}.}
  \bibinfo{year}{2017}\natexlab{}.
\newblock \showarticletitle{BlockSci: Design and applications of a blockchain
  analysis platform}.
\newblock \bibinfo{journal}{\emph{arXiv preprint arXiv:1709.02489}}
  (\bibinfo{year}{2017}).
\newblock


\bibitem[\protect\citeauthoryear{Lyu}{Lyu}{1996}]%
        {lyu96handbook}
\bibfield{author}{\bibinfo{person}{Michael~R. Lyu}.}
  \bibinfo{year}{1996}\natexlab{}.
\newblock \bibinfo{booktitle}{\emph{Handbook of Software Reliability
  Engineering}}.
\newblock \bibinfo{publisher}{McGraw-Hill, New York}.
\newblock


\bibitem[\protect\citeauthoryear{Nakamoto}{Nakamoto}{2008}]%
        {nakamoto2008bitcoin}
\bibfield{author}{\bibinfo{person}{Satoshi Nakamoto}.}
  \bibinfo{year}{2008}\natexlab{}.
\newblock \bibinfo{title}{Bitcoin: A peer-to-peer electronic cash system}.
\newblock
\newblock


\bibitem[\protect\citeauthoryear{S.}{S.}{2016}]%
        {what2}
\bibfield{author}{\bibinfo{person}{Raval S.}} \bibinfo{year}{2016}\natexlab{}.
\newblock \bibinfo{title}{Decentralized Applications: Harnessing Bitcoin's
  Blockchain Technology}.
\newblock
\newblock


\bibitem[\protect\citeauthoryear{Sarwar, Karypis, Konstan, and Riedl}{Sarwar
  et~al\mbox{.}}{2001}]%
        {sarwar:item-based}
\bibfield{author}{\bibinfo{person}{Badrul Sarwar}, \bibinfo{person}{George
  Karypis}, \bibinfo{person}{Joseph Konstan}, {and} \bibinfo{person}{John
  Riedl}.} \bibinfo{year}{2001}\natexlab{}.
\newblock \showarticletitle{Item-Based Collaborative Filtering Recommendation
  Algorithms}. In \bibinfo{booktitle}{\emph{Proc. 10th Int'l Conf. World Wide
  Web (WWW'01)}}. \bibinfo{pages}{285--295}.
\newblock


\bibitem[\protect\citeauthoryear{Silic, Delac, and Srbljic}{Silic
  et~al\mbox{.}}{2013}]%
        {DBLP:conf/sigsoft/SilicDS13}
\bibfield{author}{\bibinfo{person}{Marin Silic}, \bibinfo{person}{Goran Delac},
  {and} \bibinfo{person}{Sinisa Srbljic}.} \bibinfo{year}{2013}\natexlab{}.
\newblock \showarticletitle{Prediction of atomic web services reliability based
  on k-means clustering}. In \bibinfo{booktitle}{\emph{Joint Meeting of the
  European Software Engineering Conference and the {ACM} {SIGSOFT} Symposium on
  the Foundations of Software Engineering, ESEC/FSE'13, Saint Petersburg,
  Russian Federation, August 18-26, 2013}}. \bibinfo{pages}{70--80}.
\newblock
\urldef\tempurl%
\url{https://doi.org/10.1145/2491411.2491424}
\showDOI{\tempurl}


\bibitem[\protect\citeauthoryear{Weber, Gramoli, Ponomarev, Staples, Holz,
  Tran, and Rimba}{Weber et~al\mbox{.}}{2017}]%
        {weber2017availability}
\bibfield{author}{\bibinfo{person}{Ingo Weber}, \bibinfo{person}{Vincent
  Gramoli}, \bibinfo{person}{Alex Ponomarev}, \bibinfo{person}{Mark Staples},
  \bibinfo{person}{Ralph Holz}, \bibinfo{person}{An~Binh Tran}, {and}
  \bibinfo{person}{Paul Rimba}.} \bibinfo{year}{2017}\natexlab{}.
\newblock \showarticletitle{On availability for blockchain-based systems}. In
  \bibinfo{booktitle}{\emph{Proceedings of the 36th International Symposium on
  Reliable Distributed Systems (SRDS). IEEE}}.
\newblock


\bibitem[\protect\citeauthoryear{Wood}{Wood}{2014}]%
        {wood2014ethereum}
\bibfield{author}{\bibinfo{person}{Gavin Wood}.}
  \bibinfo{year}{2014}\natexlab{}.
\newblock \showarticletitle{Ethereum: A secure decentralised generalised
  transaction ledger}.
\newblock \bibinfo{journal}{\emph{Ethereum Project Yellow Paper}}
  \bibinfo{volume}{151} (\bibinfo{year}{2014}).
\newblock


\bibitem[\protect\citeauthoryear{Yang}{Yang}{2017}]%
        {blockguide}
\bibfield{author}{\bibinfo{person}{Baohua Yang}.}
  \bibinfo{year}{2017}\natexlab{}.
\newblock \bibinfo{booktitle}{\emph{Blockchain guide}}.
\newblock https://github.com/yeasy/.
\newblock


\bibitem[\protect\citeauthoryear{Yung-chen}{Yung-chen}{2017}]%
        {taipei}
\bibfield{author}{\bibinfo{person}{Hsieh Yung-chen}.}
  \bibinfo{year}{2017}\natexlab{}.
\newblock \bibinfo{booktitle}{\emph{Talk about Dapp Decentralized
  Application.}}
\newblock
  https://medium.com/taipei-ethereum-meetup/\%E8\%AB\%96-dapp-decentralized-application-c843e7ed2b69.
\newblock


\bibitem[\protect\citeauthoryear{Zheng, Zheng, Luo, Chen, and Liu}{Zheng
  et~al\mbox{.}}{2018}]%
        {zheng2018detailed}
\bibfield{author}{\bibinfo{person}{Peilin Zheng}, \bibinfo{person}{Zibin
  Zheng}, \bibinfo{person}{Xiapu Luo}, \bibinfo{person}{Xiangping Chen}, {and}
  \bibinfo{person}{Xuanzhe Liu}.} \bibinfo{year}{2018}\natexlab{}.
\newblock \showarticletitle{A detailed and real-time performance monitoring
  framework for blockchain systems}. In \bibinfo{booktitle}{\emph{Proceedings
  of the 40th International Conference on Software Engineering: Software
  Engineering in Practice}}. ACM, \bibinfo{pages}{134--143}.
\newblock


\bibitem[\protect\citeauthoryear{Zheng and Lyu}{Zheng and Lyu}{2010}]%
        {zheng2010collaborative}
\bibfield{author}{\bibinfo{person}{Zibin Zheng} {and}
  \bibinfo{person}{Michael~R Lyu}.} \bibinfo{year}{2010}\natexlab{}.
\newblock \showarticletitle{Collaborative reliability prediction of
  service-oriented systems}. In \bibinfo{booktitle}{\emph{Software Engineering,
  2010 ACM/IEEE 32nd International Conference on}}, Vol.~\bibinfo{volume}{1}.
  IEEE, \bibinfo{pages}{35--44}.
\newblock


\bibitem[\protect\citeauthoryear{Zheng, Xie, Dai, Chen, and Wang}{Zheng
  et~al\mbox{.}}{2017}]%
        {zheng2017overview}
\bibfield{author}{\bibinfo{person}{Zibin Zheng}, \bibinfo{person}{Shaoan Xie},
  \bibinfo{person}{Hongning Dai}, \bibinfo{person}{Xiangping Chen}, {and}
  \bibinfo{person}{Huaimin Wang}.} \bibinfo{year}{2017}\natexlab{}.
\newblock \showarticletitle{An overview of blockchain technology: Architecture,
  consensus, and future trends}. In \bibinfo{booktitle}{\emph{Big Data (BigData
  Congress), 2017 IEEE International Congress on}}. IEEE,
  \bibinfo{pages}{557--564}.
\newblock


\end{thebibliography}

\end{document}